\pdfoutput=1 
\documentclass[lettersize,journal]{IEEEtran}  

\IEEEoverridecommandlockouts                              

\usepackage[OT1]{fontenc} 
\usepackage[english]{babel}
\usepackage{cite}
\usepackage{amsmath,amssymb,amsfonts}
\usepackage{textcomp}
\usepackage{xcolor}
\usepackage{blindtext}
\usepackage{subfiles} 
\usepackage{mathtools}
\usepackage{adjustbox}
\usepackage{tikz}
\usepackage{svg}
\usepackage{booktabs} 
\usepackage[final]{microtype} 
\usepackage{algorithm} 
\usepackage[algo2e,ruled,vlined]{algorithm2e} 
\usepackage{hyperref}
\hypersetup{hidelinks, breaklinks = true}
\usepackage[capitalise]{cleveref}
\usepackage{mathdots}
\usepackage{yhmath}
\usepackage{cancel}
\usepackage{color}
\usepackage{array}
\usepackage{multirow}
\usepackage{gensymb}
\usepackage{adjustbox}
\usepackage{tabularx}
\usepackage{comment}
\usepackage{extarrows}
\usepackage{booktabs}
\usepackage{diagbox}
\usepackage{graphicx}
\usepackage{subcaption}
\usepackage[normalem]{ulem}
\useunder{\uline}{\ul}{}
\usetikzlibrary{fadings}
\usetikzlibrary{patterns}
\usetikzlibrary{shadows.blur}
\usetikzlibrary{shapes}
\graphicspath{ {./Figures/} }
\usepackage{amsmath}
\usepackage{tikz}
\usepackage{mathdots}
\usepackage{yhmath}
\usepackage{cancel}
\usepackage{color}
\usepackage{array}
\usepackage{multirow}
\usepackage{amssymb}
\usepackage{gensymb}
\usepackage{tabularx}
\usepackage{extarrows}
\usepackage{booktabs}
\usetikzlibrary{fadings}
\usetikzlibrary{patterns}
\usetikzlibrary{shadows.blur}
\usetikzlibrary{shapes}
\usepackage{algcompatible}
\SetKwComment{Comment}{/* }{ */}

\DeclareMathOperator*{\argmin}{argmin}

\title{\LARGE \bf
Cooperative Lane Changing in Mixed Traffic can be Robust to Human Driver Behavior
}

\author{Anni Li, Andres S. Chavez Armijos, and Christos G. Cassandras
\thanks{ A. Li, A. S. Chavez Armijos, and C. G. Cassandras are
with the Division of Systems Engineering and Center for Information and
Systems Engineering, Boston University, Brookline, MA 02446
(email:\{anlianni; aschavez; cgc\}@bu.edu).}
}

\begin{document}

\maketitle
\thispagestyle{empty}
\pagestyle{empty}

\begin{abstract}
We derive time and energy-optimal control policies for a Connected Autonomous Vehicle (CAV) to complete lane change maneuvers in mixed traffic. The interaction between CAVs and Human-Driven Vehicles (HDVs) requires designing the best possible response of a CAV to actions by its neighboring HDVs. This interaction is formulated using a bilevel optimization setting with an appropriate behavioral model for an HDV's.
Then, an iterated best response (IBR) method is used to determine a Nash equilibrium. However, we also show that when a common and simple-to-detect condition applies, the optimal lane-changing policy is in fact independent of HDV behavior with a CAV changing lanes by cooperating with another CAV in the target lane and always merging ahead of it. Thus, the dependence on the interaction between CAVs and HDVs may be eliminated in such cases.
Simulation results are included to show the effectiveness of our controllers in terms of cost, safety guarantees, and disruption to the traffic flow when uncontrollable HDVs are present. 

\end{abstract}


\section{INTRODUCTION}
The emergence of Connected Autonomous Vehicles (CAVs), also known as ``self-driving cars'', has the potential to significantly transform the operation of transportation networks and improve their performance by assisting drivers to make decisions so as to reduce travel times, energy consumption, air pollution, traffic congestion, and accidents. 
In highway driving, this potential manifests itself in automating lane-changing maneuvers through proper trajectory planning
\cite{luo2016dynamic} or accelerated maneuver evaluation using car-following models
\cite{zhao2017accelerated}. 
The automated lane changing problem has attracted increasing attention \cite{fisac2019hierarchical,liu2022three,duan2023cooperative}. When controlling a single vehicle, the feasibility of a maneuver depends on the state of nearby traffic \cite{kamal2012model}, and motion planning may be designed as in \cite{nilsson2016lane}. However, a lane change maneuver is often infeasible without the cooperation of other vehicles, especially under heavier traffic conditions. Several studies have addressed infeasibility issues for CAVs to perform lane change maneuvers under vehicle cooperation \cite{li2018balancing,katriniok2020nonconvex}. Moreover, cooperation among CAVs provides opportunities to perform automated lane change maneuvers both safely \cite{he2021rule} and optimally \cite{li2017optimal}. 

In our previous work, assuming a 100\% CAV penetration rate, we analytically derived cooperative joint time and energy-optimal controllers \cite{chen2020cooperative} from the point of view of a CAV executing a lane change maneuver. This ``selfish'' approach, however, may adversely affect the overall traffic throughput, a problem that was addressed in \cite{armijos2022sequential} by seeking to improve the performance of the whole traffic network in terms of both maximal throughput and minimal average maneuver time. 

However, 100\% CAV penetration is not likely in the near future, raising the question of how to benefit from the presence of at least some CAVs in mixed traffic where CAVs must interact with Human-Driven Vehicles (HDVs).
This is a challenging task that has become the focus of recent research.
For example, adaptive cruise controllers have been developed in mixed traffic environments with platoon formulations for CAVs in \cite{zheng2017platooning}, while car-following models are implemented to have a deterministic quantification of HDV states in \cite{zhao2018optimal}. In an effort to accurately model human driver behavior, 
\cite{schwarting2019social} defines the concept of social value orientation for autonomous driving to quantify an agent's degree of prosocialness or individualism and applies a game-theoretic formulation to predict human behavior. Vehicle interactions are considered in \cite{burger2022interaction,wang2019game} by using bilevel optimization to assist autonomous vehicles to apply the best possible response to an opponent's action. Towards the same goal, learning-based techniques are used in \cite{guo2020inverse,le2022cooperative}.

\begin{figure} [pt]
    \centering
    \begin{adjustbox}{width=7cm, height = 3cm,center}

\tikzset{every picture/.style={line width=0.75pt}} 

\begin{tikzpicture}[x=0.75pt,y=0.75pt,yscale=-1,xscale=1]

\draw [line width=3]    (55.72,35.16) -- (334.76,35.7) -- (583.81,35.7) ;
\draw [line width=3]    (55.72,239.03) -- (297.62,239.03) -- (583.81,239.03) ;
\draw [color={rgb, 255:red, 248; green, 231; blue, 28 }  ,draw opacity=1 ][fill={rgb, 255:red, 248; green, 231; blue, 28 }  ,fill opacity=1 ][line width=3]  [dash pattern={on 7.88pt off 4.5pt}]  (57.43,138.11) -- (583.81,136.69) ;
\draw (143.09,190.63) node  {\includegraphics[width=91.22pt,height=57.34pt]{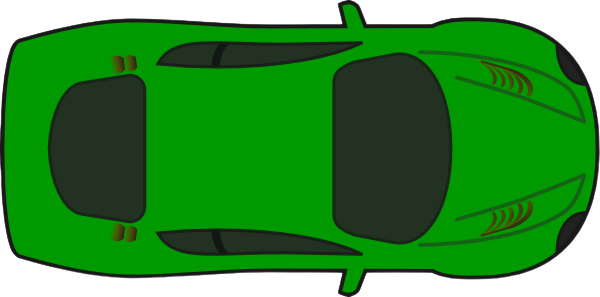}};
\draw (256.97,87.5) node  {\includegraphics[width=131.6pt,height=115.95pt]{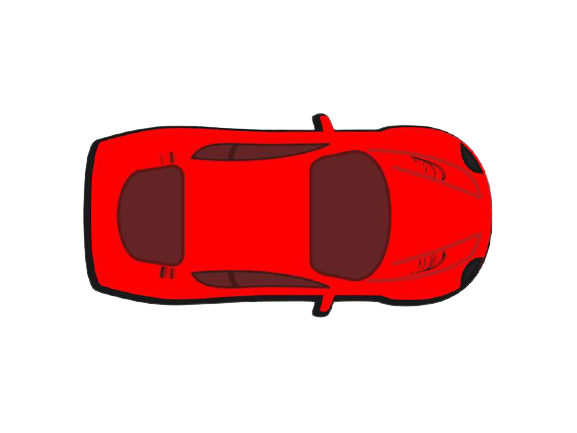}};
\draw  [dash pattern={on 4.5pt off 4.5pt}]  (211,191.47) .. controls (261.75,188.48) and (326.35,177.58) .. (354.58,94.72) ;
\draw [shift={(355,93.47)}, rotate = 108.43] [color={rgb, 255:red, 0; green, 0; blue, 0 }  ][line width=0.75]    (10.93,-3.29) .. controls (6.95,-1.4) and (3.31,-0.3) .. (0,0) .. controls (3.31,0.3) and (6.95,1.4) .. (10.93,3.29)   ;
\draw (451.09,87.24) node  {\includegraphics[width=91.22pt,height=57.34pt]{plots/greenVehicleTopView.png}};
\draw [color={rgb, 255:red, 74; green, 144; blue, 226 }  ,draw opacity=1 ] [dash pattern={on 4.5pt off 4.5pt}]  (213.5,199.47) .. controls (264.25,196.48) and (531.81,192.51) .. (564.03,95.93) ;
\draw [shift={(564.5,94.47)}, rotate = 107.02] [color={rgb, 255:red, 74; green, 144; blue, 226 }  ,draw opacity=1 ][line width=0.75]    (10.93,-3.29) .. controls (6.95,-1.4) and (3.31,-0.3) .. (0,0) .. controls (3.31,0.3) and (6.95,1.4) .. (10.93,3.29)   ;

\draw (235.18,222.16) node  [font=\large] [align=left] {\begin{minipage}[lt]{46pt}\setlength\topsep{0pt}
CAV C
\end{minipage}};
\draw (548.98,74.35) node  [font=\large] [align=left] {\begin{minipage}[lt]{40.91pt}\setlength\topsep{0pt}
CAV 1
\end{minipage}};
\draw (167.42,74.35) node  [font=\large] [align=left] {\begin{minipage}[lt]{26.64pt}\setlength\topsep{0pt}
HDV
\end{minipage}};
\draw (265,155.4) node [anchor=north west][inner sep=0.75pt]    {$( 1)$};
\draw (476,165.4) node [anchor=north west][inner sep=0.75pt]    {$( 2)$};

\end{tikzpicture}

    \end{adjustbox}
    \caption{The basic lane-changing maneuver process.}
    \label{fig:lane_change_process}
    \vspace*{-\baselineskip}
    \vspace*{-3mm}
\end{figure}

In this paper, we consider the joint time and energy-optimal automated lane change problem in the presence of mixed traffic, while at the same time limiting the overall traffic throughput disruption. 
As shown in \cite{armijos2022sequential}, a key step in this problem is to determine the optimal pair of vehicles in the fast lane that the lane-changing CAV can move in between, as shown in Fig. \ref{fig:lane_change_process}. When the red vehicle is also a CAV, this triplet can effectively cooperate leading to significant performance improvements over a baseline of 100\% HDVs. Clearly, such cooperation cannot be guaranteed when the red vehicle is an HDV in Fig. \ref{fig:lane_change_process}, therefore 
minimizing travel time, energy consumption and traffic disruption can no longer be ensured.
The contribution of this paper 
is the computation of optimal lane change trajectories for vehicle $C$ in Fig. \ref{fig:lane_change_process}
along the longitudinal traffic direction in a mixed traffic setting where the two CAVs in the figure must interact with the HDV.
We limit ourselves to the triplets shown since they provide an opportunity for two CAVs to cooperate while also interacting with the HDV
(if the relative position between HDV and CAV 1 is reversed, the problem is much simpler, while if both fast lane vehicles are HDVs the problem is harder and the subject of ongoing research).
For CAV $C$ to safely merge ahead of the HDV, it must account for this driver's behavior since the HDV is otherwise uncontrollable. However, another option is for CAV $C$ to merge ahead of the cooperating CAV 1, in which case the HDV is constrained to merely ``follow'' CAV 1. In the former case, a game-theoretic framework is established for the interactive decision-making process between the CAVs and the HDV. We use bilevel optimization to formulate this interaction in which the behavior of the HDV is estimated and considered as a constraint in the two optimization problems for the two CAVs. The latter case requires the cooperation of the CAVs and is robust to the HDV behavior which, therefore, becomes irrelevant, while safety can still be guaranteed for all vehicles involved. We derive optimal controllers for CAVs 1 and 2 in both cases which can then be compared to select the optimal one in the sense of minimizing an appropriate cost function. Moreover, we show that this optimal binary decision boils down to exceeding or not a \emph{threshold on the distance between the HDV and CAV 1} at the start of the maneuver. Intuitively, when this distance is small, it is optimal for CAV $C$ to simply merge ahead of CAV 1; conversely, when the distance is large, CAV $C$ has adequate space to position itself between the HDV and CAV 1 without causing any disruption to the HDV, hence also all traffic that follows it.

The rest of the paper is organized as follows. Section \ref{secII:ProblemFormulation} presents the formulation of the
vehicle dynamics and problem constraints. 
In Sections \ref{secIII:gametheoreticPlanning} and \ref{secIV} respectively,
complete optimal control solutions are provided for the policy of merging ahead of the HDV (by 
solving a bilevel optimization problem) and for merging ahead of the cooperating CAV 1. Section \ref{secIV:Simulation} provides simulation results for
several representative examples and we conclude with Section \ref{secV:Conclusions}.

\section{PROBLEM FORMULATION}
\label{secII:ProblemFormulation}
The lane change maneuver is triggered by CAV $C$ when an obstacle (e.g.,. slow-moving vehicle) ahead is detected or at any arbitrary time set by the CAV.
We aim to minimize the maneuver time and energy expended, while alleviating any disruption to the fast lane traffic. Moreover, considering the presence of HDVs, $C$ also needs to be aware of the behavior of its surrounding HDVs in order to guarantee safety, which requires estimating and predicting the HDV's behavior. 


For every vehicle in Fig. \ref{fig:lane_change_process}, indexed by $i=1,C,H$, its dynamics take the form
\begin{equation}
\label{eq:vehicle_dynamics}
\dot{x}_i(t) = v_i(t), ~~~\dot{v}_i(t) = u_i(t)
\end{equation}
where $x_i(t)$ is the current longitudinal position measured with respect to a given origin, $v_i(t)$ and $u_i(t)$ are the speed and (controllable) acceleration of vehicle $i$ at time $t$, respectively. The actions of vehicles $1, C, H$ are initiated at time $t_0$, where $x_C(t_0)$ is the initial position of CAV $C$, and $t_f$ is the terminal time when the longitudinal maneuver is completed. 
In this paper, we do not include the lateral component of the lane change maneuver, in which $C$ solves a decentralized optimal control
problem seeking to jointly minimize the time and energy
consumed, since this is no different than the one presented in \cite{chen2020cooperative}.
The control input and speed for all vehicles are constrained as follows:
\begin{equation*}
    u_{i_{\min}}\leq u_i(t)\leq u_{i_{\max}}, \; \forall t\in[t_0,t_f]
\end{equation*}
\begin{equation}
\label{eq:uv_constraints}
    v_{i_{\min}}\leq v_i(t)\leq v_{i_{\max}}, \; \forall t\in[t_0,t_f]
\end{equation}
where $u_{i_{\min}}<0$ and $u_{i_{\max}}>0$ denote the minimum and maximum acceleration for vehicle $i$, $v_{i_{\min}}>0$ and $v_{i_{\max}}>0$ are vehicle $i$'s allowable minimum and maximum speed, which are determined by given traffic rules. 

\textbf{Safety Constraints.} Let $d_i(v_i(t))$ be the minimum speed-dependent safe distance of vehicle $i$ with respect to its immediately preceding vehicle in the same lane:
\begin{equation}
\label{eq:safety_distance}
    d_i(v_i(t))=\varphi v_i(t)+\delta
\end{equation}
where $\varphi$ is the reaction time (generally adopted as $\varphi=1.8s$ \cite{vogel2003comparison}), $\delta$ is a constant, and $d_i(v_i(t))$ is specified from the center of vehicle $i$ to the center of its preceding vehicle. All vehicles $i=1,C,H$ in Fig. \ref{fig:lane_change_process}, must satisfy the following constraints to guarantee safety during any lane change maneuver:
\begin{subequations}
    \begin{align}
        x_1(t)-x_H(t)&\geq d_H(v_H(t)), \; \; \;\forall t\in\lbrack t_0,t_{f}] \label{eq1:cav1_hdv2_safety_constraint}\\
        x_{C}(t_{f})-x_H(t_{f})&\geq d_H(v_H(t_{f})), \label{eq2:cavC_hdv2_safety_constraint}\\
        x_1(t_{f})-x_{C}(t_{f})&\geq d_{C}(v_{C}(t_{f})). \label{eq3:cav1_cavC_safety_constraint}
     \end{align}
    \label{eq:safety_constraint_individual}
    \vspace*{-\baselineskip}
\end{subequations}
\\
where (\ref{eq1:cav1_hdv2_safety_constraint}) is the rear-end safety constraint between CAV 1 and the HDV for all $t\in[t_0,t_f]$, whereas (\ref{eq2:cavC_hdv2_safety_constraint}),(\ref{eq3:cav1_cavC_safety_constraint}) provide safety guarantees needed only at the terminal time $t_f$.

\textbf{Traffic Disruption.} 
We adopt the disruption metric introduced in \cite{armijos2022cooperative} which includes both a position and a speed disruption each measured relative to its corresponding value under no maneuver. In particular, for any vehicle $i$, the position disruption $d_x^i$, speed disruption $d_v^i$, and total disruption $D_i(t)$ at time $t$ are given by
\begin{subequations}
 \begin{align}
    &d^i_{x}(t)= 
    \begin{cases}
        \left( x_i(t)-\Bar{x}_i(t)\right )^2, &{\text{if}}\ {x_i(t)<\Bar{x}_i(t)}\\
        0, &{\text{otherwise.}}
    \end{cases}\\
    &d^i_v(t) = (v_i(t)-v_{d,i})^2 \\
    \label{eq:totaldisruption}
    &D_i(t) = \gamma_x d_x^i(t) + \gamma_v d_v^i(t)
\end{align}
\end{subequations}
where $\Bar{x}_i(t)=x_i(t_0)+v_i(t_0)(t-t_0)$ is the position of $i$ when it maintains a constant speed $v_i(t_0)$ and 
$v_{d,i}\leq v_{\max}$ is the desired speed of vehicle $i$ which matches the fast lane traffic flow.
The weights 
$\gamma_x,\gamma_v$ are selected to form a convex combination emphasizing one or the other term to reflect the total disruption generated by vehicle $i$. 

Referring to Fig. \ref{fig:lane_change_process}, we assume that CAV $C$ has already determined that it will overtake the HDV and perform the lane change either ahead of it or ahead of CAV 1. In either case, CAVs $C$ and 1 can cooperate so that the maneuver time is minimized while each CAV also minimizes its energy consumption and the disruption caused to the HDV (hence, all traffic behind it, if any). In the next two sections, each of these two decisions by CAV $C$ is analyzed and the optimal trajectories are designed. By comparing the overall costs resulting from each decision, we may then determine the optimal one. We note that the latter maneuver can be executed \emph{without any knowledge of the HDV behavior}; the only possible effect such a maneuver has on the HDV is causing some disruption if HDV has to decelerate to maintain a safe distance from CAV1.


\section{CAV C MERGES AHEAD OF HDV}
\label{secIII:gametheoreticPlanning}
\begin{figure} [pt]
    \centering
    \begin{adjustbox}{width=7cm, height = 3cm,center}

\tikzset{every picture/.style={line width=0.75pt}} 

\begin{tikzpicture}[x=0.75pt,y=0.75pt,yscale=-1,xscale=1]

\draw [line width=3]    (50.81,39.52) -- (239.1,39.21) -- (419.5,39) ;
\draw [line width=3]    (50.43,130.42) -- (218.61,131.02) -- (419.5,131) ;
\draw [color={rgb, 255:red, 248; green, 231; blue, 28 }  ,draw opacity=1 ][fill={rgb, 255:red, 248; green, 231; blue, 28 }  ,fill opacity=1 ][line width=3]  [dash pattern={on 11.25pt off 9.75pt}]  (51.33,89.28) -- (417.5,91) ;
\draw (84.9,107.49) node  {\includegraphics[width=27.15pt,height=15.76pt]{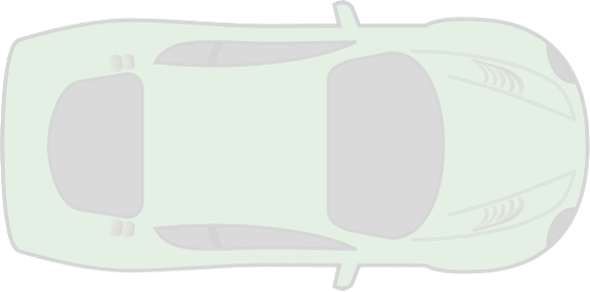}};
\draw (125.8,61.96) node  {\includegraphics[width=39.3pt,height=35.34pt]{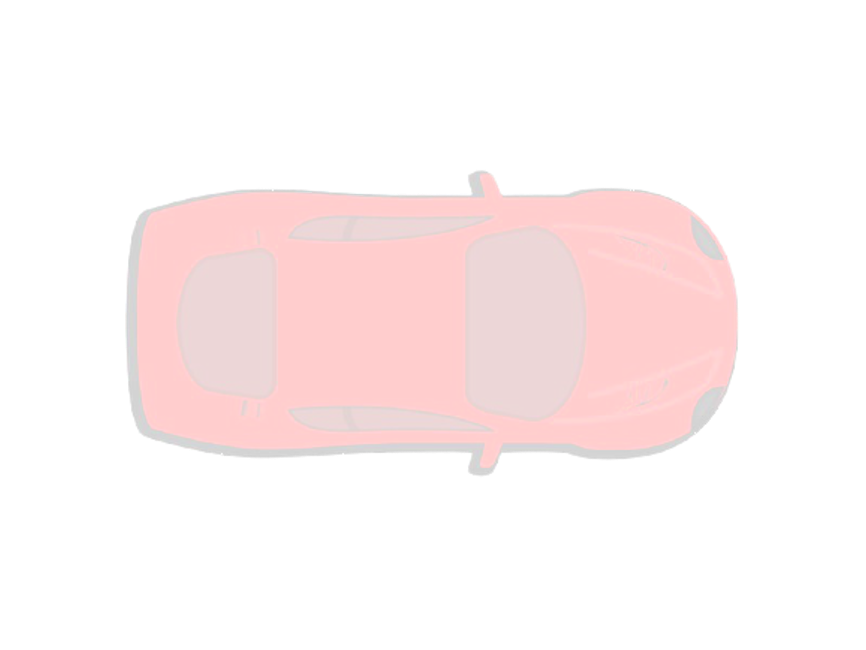}};
\draw (185.9,61.49) node  {\includegraphics[width=27.15pt,height=15.76pt]{plots/greenvehicle_t0.png}};
\draw  [color={rgb, 255:red, 65; green, 117; blue, 5 }  ,draw opacity=1 ][dash pattern={on 3.75pt off 3pt on 7.5pt off 1.5pt}] (59.82,63.44) .. controls (59.82,55.31) and (66.42,48.72) .. (74.55,48.72) -- (206.28,48.72) .. controls (214.41,48.72) and (221,55.31) .. (221,63.44) -- (221,107.61) .. controls (221,115.74) and (214.41,122.33) .. (206.28,122.33) -- (74.55,122.33) .. controls (66.42,122.33) and (59.82,115.74) .. (59.82,107.61) -- cycle ;
\draw [color={rgb, 255:red, 155; green, 155; blue, 155 }  ,draw opacity=1 ] [dash pattern={on 3.75pt off 3pt on 7.5pt off 1.5pt}]  (82.55,105.25) -- (82.53,139.52) ;
\draw [color={rgb, 255:red, 155; green, 155; blue, 155 }  ,draw opacity=1 ] [dash pattern={on 3.75pt off 3pt on 7.5pt off 1.5pt}]  (103.74,139.53) -- (122.96,139.54) ;
\draw [shift={(124.96,139.54)}, rotate = 180.04] [color={rgb, 255:red, 155; green, 155; blue, 155 }  ,draw opacity=1 ][line width=0.75]    (10.93,-3.29) .. controls (6.95,-1.4) and (3.31,-0.3) .. (0,0) .. controls (3.31,0.3) and (6.95,1.4) .. (10.93,3.29)   ;
\draw [color={rgb, 255:red, 155; green, 155; blue, 155 }  ,draw opacity=1 ] [dash pattern={on 3.75pt off 3pt on 7.5pt off 1.5pt}]  (103.71,139.53) -- (84.37,139.52) ;
\draw [shift={(82.37,139.52)}, rotate = 0.04] [color={rgb, 255:red, 155; green, 155; blue, 155 }  ,draw opacity=1 ][line width=0.75]    (10.93,-3.29) .. controls (6.95,-1.4) and (3.31,-0.3) .. (0,0) .. controls (3.31,0.3) and (6.95,1.4) .. (10.93,3.29)   ;
\draw [color={rgb, 255:red, 155; green, 155; blue, 155 }  ,draw opacity=1 ] [dash pattern={on 3.75pt off 3pt on 7.5pt off 1.5pt}]  (125.02,71.71) -- (124.96,139.54) ;

\draw (279.42,62.4) node  {\includegraphics[width=42.87pt,height=34.69pt]{plots/redVehicleTopView.png}};
\draw (359.46,62.65) node  {\includegraphics[width=32.67pt,height=18.5pt]{plots/greenVehicleTopView.png}};
\draw (279.46,105.65) node  {\includegraphics[width=32.67pt,height=18.5pt]{plots/greenVehicleTopView.png}};
\draw  [color={rgb, 255:red, 65; green, 117; blue, 5 }  ,draw opacity=1 ][dash pattern={on 3.75pt off 3pt on 7.5pt off 1.5pt}] (240.75,61.71) .. controls (240.75,53.25) and (247.61,46.39) .. (256.07,46.39) -- (392.68,46.39) .. controls (401.14,46.39) and (408,53.25) .. (408,61.71) -- (408,107.68) .. controls (408,116.14) and (401.14,123) .. (392.68,123) -- (256.07,123) .. controls (247.61,123) and (240.75,116.14) .. (240.75,107.68) -- cycle ;
\draw  [dash pattern={on 0.84pt off 2.51pt}]  (274.78,141.81) -- (274.78,29.24) ;
\draw [shift={(274.78,29.24)}, rotate = 90] [color={rgb, 255:red, 0; green, 0; blue, 0 }  ][line width=0.75]    (0,5.59) -- (0,-5.59)   ;
\draw [shift={(274.78,141.81)}, rotate = 90] [color={rgb, 255:red, 0; green, 0; blue, 0 }  ][line width=0.75]    (0,5.59) -- (0,-5.59)   ;

\draw (248.98,108.48) node [anchor=north west][inner sep=0.75pt]  [rotate=-0.04] [align=left] {$ $};
\draw (76.99,103.42) node [anchor=north west][inner sep=0.75pt]  [font=\footnotesize] [align=left] {C};
\draw (121.24,55.22) node [anchor=north west][inner sep=0.75pt]  [font=\footnotesize] [align=left] {H};
\draw (189.41,60.81) node  [font=\footnotesize] [align=left] {\begin{minipage}[lt]{10.81pt}\setlength\topsep{0pt}
1
\end{minipage}};
\draw (65.48,146.17) node [anchor=north west][inner sep=0.75pt]  [font=\footnotesize]  {$x_{2}( t_{0}) -x_{C}( t_{0})$};
\draw (268.99,101.42) node [anchor=north west][inner sep=0.75pt]  [font=\footnotesize] [align=left] {C};
\draw (271.24,57.22) node [anchor=north west][inner sep=0.75pt]  [font=\footnotesize] [align=left] {H};
\draw (356.41,63.03) node  [font=\footnotesize] [align=left] {\begin{minipage}[lt]{10.81pt}\setlength\topsep{0pt}
1
\end{minipage}};
\draw (235.76,145.17) node [anchor=north west][inner sep=0.75pt]  [font=\footnotesize]  {$x_{2}( t_{1}) =x_{C}( t_{1})$};
\draw (162.82,96.53) node [anchor=north west][inner sep=0.75pt]  [font=\normalsize] [align=left] {$\displaystyle t_{0}$$ $};
\draw (345.77,95.53) node [anchor=north west][inner sep=0.75pt]  [font=\normalsize] [align=left] {$\displaystyle t_{1}$$ $};

\end{tikzpicture}

    \end{adjustbox}
    \caption{The relative position of triplet from $t_0$ to $t_1$}
    \label{fig:relative_position_t0_t1}
    \vspace*{-\baselineskip}
    \vspace*{-3mm}
\end{figure}

Let us assume that at the start of the maneuver $t_0$, we have $x_C(t_0)<x_H(t_0)<x_1(t_0)$. Thus, we begin by separating the maneuver into two phases, 
$[t_0,t_1)$ and $[t_1,t_f]$, where $t_1$ is defined as
\begin{equation}
\label{eq:define_t1}
    t_1 = \min\{t\;|\;t \ge t_0, ~x_H(t) \leq x_C(t)\}
\end{equation}
Specifically, $t_1$ denotes the first time instant that the HDV considers any possible reaction to CAV $C$ (if $x_C(t_0)\ge x_H(t_0)$, then $t_1=t_0$). In other words, there is no interaction between CAV $C$ and the HDV until $t_1$. 
The relative position of the triplet over the two phases is shown in Fig. \ref{fig:relative_position_t0_t1}. 
In Phase I, CAV $C$ plans a trajectory which jointly minimizes $t_1$ and its energy consumption over $[t_0,t_1)$.
In Phase II, CAV $C$ estimates the behavior of the HDV and solves a bilevel optimization problem leading to a solution based on an iterated best response (IBR) algorithm.

\subsection{Optimal Trajectory for CAV C in Phase I}

Assuming that CAV $1$ and HDV travel with constant speed in Phase I, CAV $C$ can solve the following optimal control problem termed $\textbf{OCP}_{[t_0,t_1]}$:
\begin{subequations}
    \begin{align}
    \label{eq:OCP_cavC_cost}   J_{C,1}^{I}=\min\limits_{t_1,u_C(t)} \int_{t_0}^{t_1} [\alpha_{t}&+\frac{\alpha_{u}}{2}u_C^2(t)]dt + \alpha_{v} (v_C(t_1)-v_{d,1})^2 \\
    \nonumber
    s.t. \; \; &(\ref{eq:vehicle_dynamics}),(\ref{eq:uv_constraints})\\
    \label{eq1:OCP_cavC_position}
    & x_C(t_1) = x_H(t_1) \\
    \label{eq2:OCP_cavC_time}
    &t_0 \leq t_1 \leq T
    \end{align}
    \label{eq:OCP_cavC}
\end{subequations}
The cost (\ref{eq:OCP_cavC_cost}) combines the travel time $t_1-t_0$ and an energy term $u_C^2(t)$ along with a terminal cost on the speed $v_C(t_1)$, where $\alpha_{\{u,t,v\}}$ are adjustable non-negative [properly normalized weights. Constraint (\ref{eq1:OCP_cavC_position}) follows from the definition of $t_1$, and (\ref{eq2:OCP_cavC_time}) gives a maximum allowable time $T$ for CAV $C$ to perform lane change maneuvers. If (\ref{eq2:OCP_cavC_time}) is violated, the maneuver is aborted at $t_0$. 

However, $\textbf{OCP}_{[t_0,t_1]}$ may be infeasible if the initial states are such that $v_H(t_0)>v_C(t_0)$, $x_H(t_0)>x_C(t_0)$ and the allowable maneuver time $T$ is small. To allow for such possible infeasibility, we consider two additional policies that CAV $C$ can adopt. The first is to simply speed up with a constant acceleration $u_{i_{\max}}$ so that
\begin{gather}
\label{eq:constant_acc_profile}
    u_C(t)= 
    \begin{cases}
        u_{C_{\max}}, &\forall t\in [t_0,\frac{v_{C_{\max}}-v_C(t_0)}{u_{C_{\max}}}]\\
        0, &\forall t\in [\frac{v_{C_{\max}}-v_C(t_0)}{u_{C_{\max}}},t_1]
    \end{cases}
\end{gather}
which allows for the possibility that the maximum speed $v_{C_{\max}}$ is achieved before $t_1$, which is obtained from
$x_C(t_1)=x_H(t_0)+v_H(t_0)(t_1-t_0)$. 
Using the same cost function as (\ref{eq:OCP_cavC_cost}) with $u_C(t)$ in (\ref{eq:constant_acc_profile}) we obtain the cost $J_{C,2}^{I}$ for this constant acceleration policy. 

The second alternative policy exploits the cooperation capabilities between CAVs, so that CAV 1 may decelerate to induce a deceleration of HDV due to the safety requirement \eqref{eq1:cav1_hdv2_safety_constraint}. If HDV decelerates, the time for $C$ to catch up with HDV is reduced. The resulting OCP can be formulated as
\begin{subequations}
    \begin{align}
    \nonumber
    J_{C,3}^{I} = &\min\limits_{t_1, u_1(t), u_C(t)}  \int_{t_0}^{t_1} [\frac{\alpha_{u}}{2}u_1^2(t) + \frac{\alpha_{u}}{2}u_C^2(t) + \alpha_{t}]dt \\
    \label{eq1:OCP_cav1C_cost}
    &+ \alpha_{v}[(v_C(t_1)-v_{d,C})^2 + (v_1(t_1)-v_{d,1})^2]\\ 
    \nonumber
    s.t. \; \; &(\ref{eq:vehicle_dynamics}),~(\ref{eq:uv_constraints}),~(\ref{eq2:OCP_cavC_time})\\
    \label{eq2:OCP_cav1C_position_12}
    &x_1(t_1)=x_C(t_1)+d_H(v_H(t_1))
    \end{align}
    \label{eq:OCP_cav1C}
\end{subequations}
Different from \eqref{eq:OCP_cavC_cost}, here \eqref{eq1:OCP_cav1C_cost} minimizes travel time, energy, and speed disruption for both CAVs 1 and $C$. 
The constraint (\ref{eq2:OCP_cav1C_position_12}) ensures that the rear-end safety constraint (\ref{eq1:cav1_hdv2_safety_constraint}) is activated by CAV 1's action. Note that 
(\ref{eq1:OCP_cavC_position}) is used in (\ref{eq2:OCP_cav1C_position_12}) to eliminate any dependence on $x_H(t)$ and we set $d_H(v_H(t_1))=\varphi v_H(t_0)+\delta$.
Thus, the third cost $J_{C,3}^{I}$ is obtained.

The solution to OCPs (\ref{eq:OCP_cavC}) and (\ref{eq:OCP_cav1C}) can be analytically obtained through standard Hamiltonian analysis similar to OCPs formulated and solved in \cite{chen2020cooperative}. Thus, we omit the details.
In summary, the non-cooperative OCP (\ref{eq:OCP_cavC}), constant acceleration formulation (\ref{eq:constant_acc_profile}), and cooperative OCP (\ref{eq:OCP_cav1C}) provide three different control policies for CAV $C$ and we can select the optimal one through
\begin{equation}
\label{eq:pick_min_cost}
    J_C^{I} = \min \{J_{C,1}^{I},J_{C,2}^{I},J_{C,3}^{I} \},
\end{equation}
Consequently, we can also determine the optimal time $t_1^*$ marking the end of Phase I for CAV $C$.


          

\subsection{Optimal Trajectory for CAV C in Phase II}
\begin{figure}[hpbt]
    \centering   
    \includegraphics[width=\linewidth, height=5.5cm]{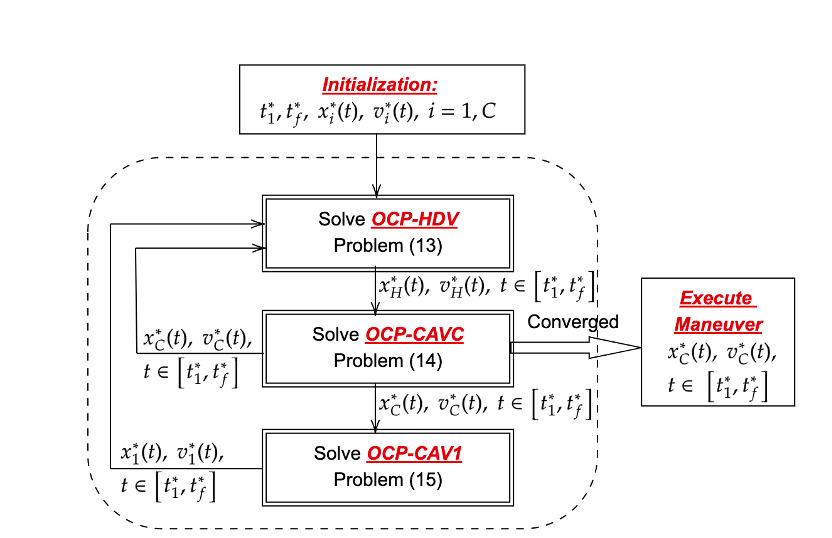} 
    
    \caption{Bilevel optimization problem solved by CAV $C$. Initialization consists of solving $\textbf{OCP}_{[t_0,t_1]}$ to obtain $t_1^*$ and $\textbf{OCP}_{[t_1^*,t_f]}$ to obtain 
    $t_f^*,x_C^*(t),v_C^*(t)$. In addition, 
    $x_1^*(t)=x_1(t_1^*)+v_1(t_1^*)(t_f^*-t_1^*),v_1^*(t)=v_1(t_1^*)$. Upon convergence, the lane change maneuver is executed with the final $x_C^*(t),v_C^*(t),t\in[t_1^*,t_f^*]$.}
    \label{fig:bilevel_diagram}
\end{figure}

The ideal optimal trajectory for CAV $C$ in Phase II in order to merge ahead of the HDV 
is obtained by an OCP we term $\textbf{OCP}_{[t_1^*,t_f]}$,
since it shares the same cost function as $\textbf{OCP}_{[t_0,t_1]}$ in (\ref{eq:OCP_cavC_cost}) except for the new time interval. It also shares the vehicle dynamics (\ref{eq:vehicle_dynamics}), speed and control limits (\ref{eq:uv_constraints}), 
and (\ref{eq2:OCP_cavC_time}) which becomes $t_1^* \leq t_f \leq T$. It differs only in the terminal state constraint which is now the rear-end safety requirement:
\begin{equation}
\label{eq2:ocp_cavC_tf_position}
    x_C(t_f)\geq x_H(t_1^*)+v_H(t_1^*)(t_f-t_1^*)+d_H(v_H(t_1^*))
\end{equation}
The solution is ``ideal'' because it assumes the HDV travels at constant speed in (\ref{eq2:ocp_cavC_tf_position}), hence ignoring any reaction that the human driver might have when detecting the lane change action of CAV $C$.
In reality, for $C$ to complete this maneuver safety and optimally, it has to estimate 
the behavior of $H$ and adjust its own trajectory based on $H$'s response. Similarly, $H$ then needs to adjust its trajectory by reacting to $C$'s response. In order to model this process,
we formulate a bilevel optimization problem for each $i=1,C,H$ in the following three subsections. We emphasize that this problem is solved by CAV $C$ and describe its structure in Fig. \ref{fig:bilevel_diagram}. 

\subsubsection{Estimate HDV Trajectory (\textbf{OCP-HDV})}
\begin{figure}[hpbt]
    \centering   
    \includegraphics[width=\linewidth, height=3.5cm]{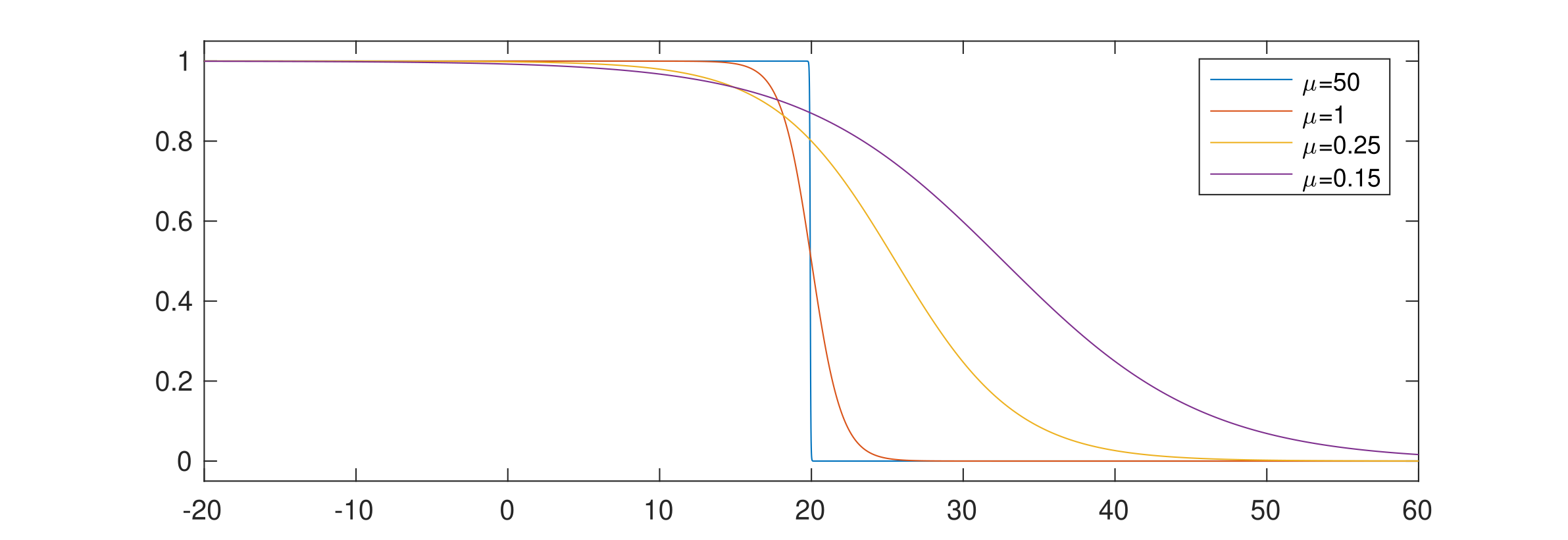} 
    
    \caption{Sigmoid safety functions, $d=20$}
    \label{fig:safefunction}
\end{figure}

We estimate the trajectory of an HDV by assuming that a human driver considers three factors: $(i)$ maintaining a constant speed that minimally deviates from some desired value $v_{d,H}$, $(ii)$ if it needs to change speeds, it does so by minimizing its acceleration/deceleration, which also saves fuel, $(iii)$ guaranteeing its safety (collision avoidance). 
To model the latter, we define a \emph{safety function} $s(\cdot)$ as a decreasing function in $x_C(t)-x_H(t)$ since a closer distance between $H$ and $C$ corresponds to a higher collision risk. We adopt the sigmoid function:
\begin{equation}
\small
    \label{eq:safety_cost}
    s(x_C(t)-x_H(t))=\frac{1}{1+\mu \exp\left(\mu(x_C(t)-x_H(t)-d)\right)}
\end{equation}
where $\mu$ is adjustable to capture different unsafe regions for different drivers. The effect of $\mu$ is shown in Fig. \ref{fig:safefunction}. One can also adjust $d$ to define the size of the unsafe region. 

We can now formulate $\textbf{OCP-HDV}$ as the problem whose solution is the estimated trajectory that CAV $C$ uses in adjusting its own response by updating $u_C(t)$: 
\begin{subequations}
    \begin{align}
    \nonumber
        \min\limits_{u_H(t)} &\int_{t_1^*}^{t_f^*} [\frac{\beta_u}{2}u_H^2(t)+\beta_v(v_H(t)-v_{d,H})^2 \\
        \label{eq:hdv_ocp_cost} &\;\;\;\;\;\;\;\;\;\;\;\;\;\;\;\;\;\;\;\;\;+\beta_s s(x_C^*(t)-x_H(t))] dt\\
        \nonumber
        s.t. \; \; &(\ref{eq:vehicle_dynamics}),~(\ref{eq:uv_constraints})\\ 
        \label{eq1:hdv_ocp_safety12}
         &x_1^*(t)-x_H(t)\geq d_H(v_H(t)),\; \forall t\in[t_1^*,t_f^*]
    \end{align}
    \label{eq:hdv_ocp}
\end{subequations}
where $\beta_{\{u,v,s\}}$ are the non-negative appropriately normalized weights that describe the characteristics of the HDV, i.e., the behavior of the driver. Constraint
(\ref{eq1:hdv_ocp_safety12}) denotes the safety constraint between the HDV and its current preceding vehicle CAV 1 for all $t\in[t_1^*,t_f^*]$. 
We immediately note that $x_C^*(t)$ and $x_1^*(t)$ are unknown to the HDV (except in the first iteration in Fig. \ref{fig:bilevel_diagram} where the initial ``ideal'' trajectories are used).
In fact, these are determined by the two lower-level problems (\ref{eq:ibr_ocp_cavC}) and (\ref{eq:cav1_ocp}) defined next, in response to the HDV's behavior expressed through $x_H^*(t),v_H^*(t)$, $t \in [t_1^*, t_f^*]$ from (\ref{eq:hdv_ocp}).


\subsubsection{Update CAV $C$ Trajectory (\textbf{OCP-CAVC})}
Similar to \textbf{OCP-HDV}, we formulate a bilevel optimization problem $\textbf{OCP-CAVC}$ for CAV $C$:
\begin{subequations}
    \begin{align}
    \label{eq:ibr_ocp_cavC_cost}
    \min\limits_{u_C(t)} \int_{t_1^*}^{t_f^*} &\frac{\alpha_u}{2}u_C^2dt+ \alpha_v(v_C(t_f^*)-v_{d,C})^2\\
        \nonumber
        s.t. \; \; &(\ref{eq:vehicle_dynamics})(\ref{eq:uv_constraints})\\ 
        \label{eq1:ibr_ocp_cavC_position}
        & x_C(t_f^*)\geq x_H^*(t_f^*)+ d_H(v_H^*(t_f^*))
    \end{align}
    \label{eq:ibr_ocp_cavC}
\end{subequations}
The position $x_H^*(t_f^*)$ in safety constraint (\ref{eq1:ibr_ocp_cavC_position}) is the optimal terminal position of $H$ given by (\ref{eq:hdv_ocp}). 
Problem (\ref{eq:ibr_ocp_cavC}) then provides the best response strategy of CAV $C$ and determines $x_C^*(t),v_C^*(t),u_C^*(t),t\in[t_1^*,t_f^*]$. Note that this information can now be provided to $\textbf{OCP-CAVC}$ as shown in Fig. \ref{fig:bilevel_diagram}.

\subsubsection{Update CAV 1 Trajectory (\textbf{OCP-CAV1})}

Since CAV 1 is cooperating with CAV $C$, CAV 1's strategy is based on the optimal policy of CAV $C$ by applying a similar bilevel optimization problem $\textbf{OCP-CAV1}$:
\begin{subequations}
    \begin{align}
    \label{eq:cav1_ocp_cost}
        \min\limits_{u_1(t)} \int_{t_1}^{t_f^*} &\frac{\alpha_u}{2}u_1^2(t)dt+\alpha_v(v_1(t_f^*)-v_d)^2\\
        \nonumber
        s.t. \; \; &(\ref{eq:vehicle_dynamics})(\ref{eq:uv_constraints})\\ 
        \label{eq:cav1_ocp_safety}
        &x_1(t_f^*)-x_C^*(t_f^*)\geq d_C(v_C^*(t_f^*)).
    \end{align}
    \label{eq:cav1_ocp}
\end{subequations}
The position $x_C^*(t_f^*)$ in safety constraint (\ref{eq:cav1_ocp_safety}) is the optimal terminal position of $C$
from $\textbf{OCP-CAVC}$. The solution of (\ref{eq:cav1_ocp}) provides the optimal trajectories $x_1^*(t),v_1^*(t),u_1^*(t),t\in[t_1^*,t_f^*]$ for CAV 1. Note that this information can now be provided to $\textbf{OCP-CAVC}$ as shown in Fig. \ref{fig:bilevel_diagram}.

\subsubsection{Iterated Best Response}
\label{sec:ibr}
The solution to each of the problems (\ref{eq:hdv_ocp}), (\ref{eq:ibr_ocp_cavC}) and (\ref{eq:cav1_ocp}) is complicated by the fact that it is coupled to the others through safety constraint or safety cost. 
Nonetheless, the problems can be jointly solved through an iterated best response (IBR) process \cite{wang2020multi} as shown in Fig. \ref{fig:bilevel_diagram} to obtain a Nash equilibrium and the corresponding optimal trajectory of CAV $C$, $x_C^*(t),v_C^*(t),t\in[t_1^*,t_f^*]$. 
This, in turn, provides the optimal cost for Phase II, $J_C^{II}$. Combining this with $J_C^{I}$ in (\ref{eq:pick_min_cost}) yields the optimal cost of the CAV $C$ policy ``merge ahead of HDV'', $J_{C,H} = J_C^{I} + J_C^{II}$. The IBR process is summarized in Algorithm \ref{alg:IBR_process}. 

Note that problems (\ref{eq:ibr_ocp_cavC}) and (\ref{eq:cav1_ocp}) can be solved analytically through standard Hamiltonian analysis as in \cite{chen2020cooperative}. The solution of (\ref{eq:hdv_ocp}) is complicated by the presence of the nonlinear safety function, but can be numerically solved. 

\begin{algorithm}[hpt]
    \caption{Iterated Best Response Process}
    \SetKwInOut{Input}{input}
    \SetKwInOut{Output}{output}
    \Input{Initial Conditions $x_i(t_1^*),v_i(t_1^*),i=1,C,H$, Relaxation Constant $\lambda$,Desired speed $v_{d,i}$, Maximum Time $T$, Iteration rounds $N$, Error Tolerance $\epsilon$.}
    \Output{$t_f^*$, Optimal Longitudinal Trajectories:\\ $x_i^*(t),v_i^*(t),u_i^*(t), t\in [t_1^*,t_f^*],i=1,C$}
    \SetKwBlock{Beginn}{beginn}{ende}
    \Begin{
        $t^*_f,x_{C,1}(t),v_{C,1}(t),t\in[t_1^*,t_f^*]\leftarrow$ Solve $\textbf{OCP}_{[t_1^*,t_f^*]}$\\
        $x_{1,1}(t)=x_1(t_1^*)+v_1(t_1^*)(t-t_1^*),\\
        v_{1,1}(t)=v_1(t_1^*),t\in[t_1^*,t_f^*]$,\\
        \While{$t_f^*\leq T$}{
        \For{$k=1$ to $N$}{
            $x_{H,k}(t_f^*),v_{H,k}(t_f^*)\leftarrow$ Solve $\textbf{OCP-HDV}$
            (\ref{eq:hdv_ocp})\\
            \If{$k\geq 2$}{
            \eIf{$||u^*_{C,k}(t)-u^*_{C,k-1}(t)||\leq \epsilon$}{
                break
            } 
            {
            $x_{C,k+1}(t),v_{C,k+1}(t),u_{C,k+1}(t),\leftarrow$ Solve \textbf{OCP-CAVC} (\ref{eq:ibr_ocp_cavC})\\
            $x_{1,k+1}(t),v_{1,k+1}(t),u_{1,k+1}(t),\leftarrow$ Solve \textbf{OCP-CAV1} (\ref{eq:cav1_ocp})\\
            $k=k+1$\\
            } 
            } 
        }
        \If{$||u^*_{C,N}(t)-u^*_{C,N-1}(t)||> \epsilon$}{
                   $ t_f^* = \lambda t_f^*$\\
                } 
    }
    }
    \label{alg:IBR_process}
\end{algorithm}

\textbf{IBR convergence}.
The convergence of IBR processes as in Fig. \ref{fig:bilevel_diagram} is generally hard to establish. However, the specific structure of the problems here facilitates such analysis. In particular, convergence depends on the initial states of the vehicles. Since the process starts with (\ref{eq:hdv_ocp}), observe that its solution depends on $x_1^*(t)$ only through the constraint 
(\ref{eq1:hdv_ocp_safety12}) and on $x_C^*(t)$ through the safety function $s(x_C(t)-x_H(t))$.
Therefore, if the distance between vehicles 1 and $H$ is larger than the minimum safety distance of the HDV, the constraint (\ref{eq1:hdv_ocp_safety12}) remains inactive and the dependence on $x_1^*(t)$ is eliminated. Similarly, 
if $C$'s speed is greater than $H$ at $t_1$, their relative distance will increase and the value of $s(x_C(t)-x_H(t))$ becomes zero, leading to the solution of (\ref{eq:hdv_ocp}) becoming independent of $x_C^*(t)$ as well, hence leading
to the convergence of the iteration process.
Conversely, if vehicles 1 and $H$ are close, or the speed of the HDV exceeds that of $C$, the dependence on $x_1^*(t)$ and $x_c^*(t)$ may not vanish, hence reducing the rate of convergence. In this case, however, as we will see next, the optimal action of CAV $C$ becomes that of merging ahead of CAV 1, thus rendering the IBR process irrelevant.

A formal convergence analysis of the IBR process has yet to be carried out. In practice, to implement the IBR process, we predetermine a number of iterations $N$ and error tolerance $\varepsilon$. If the process has not converged within $\varepsilon$ after $N$ iterations, we relax the terminal time and repeat the process. If the terminal time reaches a given threshold, we end the process and apply the final $u^*_{C}(t)$ as the CAV $C$ control.

\section{CAV C MERGES AHEAD OF CAV 1} 
\label{secIV}
In this section, we consider the alternative CAV $C$ policy to merge ahead of $CAV 1$ rather than the HDV. We immediately see that if this policy leads to an optimal cost $J_{C,1}$ such that $J_{C,1} \leq J_{C,H}$, this makes it not only optimal but also independent of the HDV behavior, since the HDV's action cannot affect CAV $C$ and the HDV is limited to maintaining a safe distance from CAV $1$.

The optimal trajectory in this case is obtained jointly with that of the cooperating CAV 1 by solving the problem: 
\begin{subequations}
    \begin{align}
    \nonumber
         \min\limits_{t_f,u_1,u_C} &\int_{t_0}^{t_f} [\frac{\alpha_u}{2}(u_1^2(t)+u_C^2(t))+\alpha_t]dt\\
         \label{eq:cav1C_ocp_cost}
         &+\frac{\alpha_v}{2} [(v_C(t_f)-v_{d,C})^2+(v_1(t_f)-v_{d,1})^2]\\
        \nonumber
        s.t. \; \; &(\ref{eq:vehicle_dynamics})(\ref{eq:uv_constraints})\\ 
        \label{eq:cav1C_ocp_safety}
        & x_C(t_f)-x_1(t_f)=d_1(v_1(t_f)).
    \end{align}
    \label{eq:cav1C_ocp}
\end{subequations}
where $\alpha_{\{t,u,v\}}$ are adjustable properly normalized weights for travel time, energy, and speed deviation, respectively. The problem (\ref{eq:cav1C_ocp}) can be analytically solved by standard Hamiltonian analysis.

Let $\mathbf{x}_i(t):=(x_i(t),v_i(t))$ and $\mathbf{\lambda}_i(t)=(\lambda_i^x(t),\lambda_i^v(t))^T$ be the state and costate vector for vehicles $i=1,C$, respectively. The Hamiltonian for (\ref{eq:cav1C_ocp}) with state constraint, control constraint adjoined is 
\begin{multline}
    \label{eq:hamiltonian} H(\mathbf{x_C},\mathbf{\lambda_C},u_C,\mathbf{x_1},\mathbf{\lambda_1},u_1)=
    \frac{1}{2}u_C^2+\frac{1}{2}u_1^2+\alpha_t\\+\lambda_C^x v_C+\lambda_C^v u_C+\lambda_1^x v_1+\lambda_1^v u_1 \\+\mu_1(v_{1_{\min}}-v_1)+
    \mu_2(v_1-v_{1_{\max}})\\+\mu_3(u_{1_{\min}}-u_1)+
    \mu_4(u_1-u_{1_{\max}})\\+\eta_1(v_{C_{\min}}-v_C)+
    \eta_2(v_C-v_{C_{\max}})\\+\eta_3(u_{C_{\min}}-u_C)+
    \eta_4(u_C-u_{C_{\max}}).
\end{multline}
The Lagrange multipliers $\mu_1,\mu_2,\mu_3,\mu_4,\eta_1,\eta_2,\eta_3,\eta_4$ are positive when their corresponding constraints are active and become 0 when the constraints are inactive. The problem has an unspecified terminal time $t_f$, and the terminal position of vehicles $1,C$ are constrained by a function $\psi:=x_C(t_f)-x_1(t_f)-\varphi v_1(t_f)-\delta=0$. Beside, the problem has a terminal cost $\phi:= \frac{\alpha_v}{2}[(v_1(t_f)-v_{d,1})^2+(v_C(t_f)-v_{d,C})^2]$. The terminal constraint and cost are not the explicit function of time. The transversality condition is given as 
\begin{equation}  
\label{appeq:transversality}
H(\mathbf{x_C},\mathbf{\lambda_C},u_C,\mathbf{x_1},\mathbf{\lambda_1},u_1)|_{t=t_f}=0,
\end{equation}
with $\mathbf{\lambda}(t_f)=(\frac{\partial \phi}{\partial \mathbf{x}}+\nu^T \frac{\partial \psi}{\partial \mathbf{x}})^T |_{t=t_f}$ as the costate boundary conditions, where $\nu$ denotes a Lagrange multiplier. The Euler-Lagrange equations become
\begin{align}
\label{eq:eular_lagrange}
\nonumber\dot{\lambda}_C^x&=-\dfrac{\partial H}{\partial x_C}= 0,\\
\nonumber
\dot{\lambda}_C^v&=-\dfrac{\partial H}{\partial v_C}=-\lambda_C^x+\eta_1-\eta_2,\\
\nonumber\dot{\lambda}_1^x&=-\dfrac{\partial H}{\partial x_1}= 0,\\
\dot{\lambda}_1^v&=-\dfrac{\partial H}{\partial v_1}=-\lambda_1^x+\mu_1-\mu_2,
\end{align}
and the necessary conditions for optimality are
\begin{align}
\nonumber
\label{appeq:optimality_condition}
&\dfrac{\partial H}{\partial u_C}=\alpha_u u_C(t)+\lambda_C^v(t)-\eta_3+\eta_4=0,\\
&\dfrac{\partial H}{\partial u_1}=\alpha_u u_1(t)+\lambda_1^v(t)-\mu_3+\mu_4=0.
\end{align}
Suppose all the constraints are inactive for $t\in[t_1,t_f]$, we have $\mu_1=\mu_2=\mu_3=\mu_4=\eta_1=\eta_2=\eta_3=\eta_4=0$. Apply the Eular-Lagrange equations in (\ref{eq:eular_lagrange}), we get $\dot{\lambda}_1^x=\dot{\lambda}_C^x=0$ and  $\dot{\lambda}_1^v=-\lambda_1^x(t),\dot{\lambda}_C^v=-\lambda_C^x(t)$ which imply that $\lambda_1^x=a_1,\lambda_C^x=a_C$ and $\lambda_1^v=-(a_1t+b_1),\lambda_C^v=-(a_Ct+b_C)$, respectively. The parameters $a_1,b_1,a_C,b_C$ here are integration constants. From the optimality condition (\ref{appeq:optimality_condition}),we have 
\begin{subequations}
\label{appeq:opt_condition}
\begin{align}
    &\alpha_u u_1(t)+\lambda_1^v=0,\\
    &\alpha_u u_C(t)+\lambda_C^v=0.
\end{align}
\end{subequations}
Consequently, we obtain the following optimal solutions
\begin{subequations}
    \begin{align}
        &u^*_1(t)=\frac{1}{\alpha_u}(a_1t+b_1),\\
        &u^*_C(t)=\frac{1}{\alpha_u}(a_Ct+b_C),\\
        &v_1^*(t)=\frac{1}{\alpha_u}(\frac{1}{2}a_1t^2+b_1t+c_1),\\
        &v_C^*(t)=\frac{1}{\alpha_u}(\frac{1}{2}a_Ct^2+b_Ct+c_C),\\
        &x_1^*(t)=\frac{1}{\alpha_u}(\frac{1}{6}a_1t^3+\frac{1}{2}b_1t^2+c_1t+d_1),\\
        &x_C^*(t)=\frac{1}{\alpha_u}(\frac{1}{6}a_Ct^3+\frac{1}{2}b_Ct^2+c_Ct+d_C),
    \end{align}
\end{subequations}
where $c_1,d_1,c_C,d_C$ are also integration constants. Moreover, considering the boundary condition of the costate vector at time $t_f$, we have $\lambda_1^x(t_f)=a_1,\lambda_C^x(t_f)=a_C$ and 
\begin{subequations}
    \begin{align}
        &\lambda_1^x(t_f)=(\frac{\partial \phi}{\partial x_1}+\nu \frac{\partial \psi}{\partial x_1})|_{t=t_f}=-\nu,\\
        &\lambda_1^v(t_f)=(\frac{\partial \phi}{\partial v_1}+\nu \frac{\partial \psi}{\partial v_1})|_{t=t_f}=\alpha_v(v_1(t_f)-v_{d,1})-\nu\varphi,\\
        &\lambda_C^x(t_f)=(\frac{\partial \phi}{\partial x_C}+\nu \frac{\partial \psi}{\partial x_C})|_{t=t_f}=\nu,\\
        &\lambda_C^v(t_f)=(\frac{\partial \phi}{\partial v_C}+\nu \frac{\partial \psi}{\partial v_C})|_{t=t_f}=\alpha_v(v_C(t_f)-v_{d,C}).
    \end{align}
\end{subequations}
The transversality condition (\ref{appeq:transversality}) gives the following relationship
\begin{align}
\label{appeq:terminal_costate}
\nonumber
    \frac{\alpha_u}{2}&u_C^2(t_f)+\frac{\alpha_u}{2}u_1^2(t_f)+\alpha_t+\lambda_C^x(t_f) v_C(t_f)\\
    &+\lambda_C^v(t_f) u_C(t_f)+\lambda_1^x(t_f) v_1(t_f)+\lambda_1^v(t_f) u_1(t_f)=0
\end{align}
Therefore, combining all the equations (\ref{appeq:opt_condition})-(\ref{appeq:terminal_costate}), we can solve the following nonlinear algebraic equations for $a_i,b_i,c_i,d_i,i=1,C$ and $t_f,\nu$:
\begin{subequations}
    \begin{align}
        &a_1 = -\nu,\\
        &a_c = \nu,\\
        &a_1 t_f+b_1=\alpha_v(v_{d,1}-v_1(t_f))+\nu\varphi,\\
        &a_C t_f+b_C=\alpha_v(v_{d,C}-v_C(t_f)),\\
        &\frac{1}{\alpha_u}(\frac{1}{2}a_1t_1^2+b_1t_1+c_1)=v_1(t_1),\\
        &\frac{1}{\alpha_u}(\frac{1}{2}a_Ct_1^2+b_Ct_1+c_C)=v_C(t_1),\\
        &\frac{1}{\alpha_u}(\frac{1}{6}a_1t_1^3+\frac{1}{2}b_1t_1^2+c_1t_1+d_1)=x_1(t_1),\\
        &\frac{1}{\alpha_u}(\frac{1}{6}a_Ct_1^3+\frac{1}{2}b_Ct_1^2+c_Ct_1+d_C)=x_C(t_1),\\
        &x_C(t_f)-x_1(t_f)=\varphi v_1(t_f)+\delta,\\
        &-\frac{1}{2}(b_C^2+a_C^2)+\alpha_u \alpha_t+(a_Cc_C+a_1c_1)=0.
    \end{align}
\end{subequations}

A solution for $t_f^*$ and $x_i^*(t),v_i^*(t),u_i^*(t),i=1,C$ for $t\in[t_0,t_f^*]$
can be analytically obtained. The corresponding cost is denoted by $J_{C,1}$. Clearly, if $J_{C,1} \leq J_{C,H}$ then CAV $C$ selects this policy which depends only on the cooperation between CAVs 1 and $C$, thus making it \emph{independent of the HDV's behavior}. 
Lastly, the HDV trajectory, in this case, is estimated using \eqref{eq:hdv_ocp} with $\beta_s=0$, since CAV $C$ would not merge ahead of the HDV.

\section{SIMULATION RESULTS}
\label{secIV:Simulation}
This section provides simulation results illustrating the time and energy optimal lane changing trajectories for each CAV in mixed traffic and illustrates when CAV C  should merge ahead of CAV 1  so as to render the maneuver independent of the HDV behavior.
Our simulation setting is that of Fig.\ref{fig:lane_change_process}. The allowable speed range is $v\in[15,35] m/s$, and the acceleration of vehicles is limited to $u\in[-7,3.3] m/s^2$. The desired speed for the CAVs is considered as the traffic flow speed, which is set to $30 m/s$. The desired speed for HDV is assumed to be the same as its initial speed. To guarantee safety, the inter-vehicle safe distance is given by $\delta = 1.5 m$, and the reaction time is $\varphi = 0.6 s.$ The disruption in (\ref{eq:totaldisruption}) is evaluated with parameters $\gamma_x=0.5,\gamma_v=0.5$. When any of the problems (\ref{eq:hdv_ocp}), (\ref{eq:ibr_ocp_cavC}), or (\ref{eq:cav1_ocp}) is infeasible or whenever the optimal trajectory of $C$ does not converge, we relax the terminal time with a relaxation rate $\lambda=1.8$. The numerical solutions to the optimization problems are obtained using an interior point optimizer (IPOPT) on an Intel(R) Core(TM) i7-8700 3.20GHz.

\textbf{``Merge ahead of HDV'' policy.}
As discussed in Section III, in order for CAV $C$ to evaluate the cost of this policy, it breaks down its trajectory into two phases if its initial position is behind the HDV. 
Thus, in Phase I, we solve problems (\ref{eq:OCP_cavC}), (\ref{eq:OCP_cav1C}), and (\ref{eq:constant_acc_profile}) to obtain the minimum cost, hence the optimal trajectory for Phase I.
The weights $\alpha_{\{t,u,v\}}$ in \eqref{eq:OCP_cavC} and \eqref{eq:OCP_cav1C} are set to 0.55, 0.2, and 0.25, respectively. The maximum maneuver time is set as $T=15s$. If any of the OCPs is infeasible in this phase, its corresponding cost is set as ``$Inf$''. The results are shown in Table \ref{tab:vehicleCSample}, where we see that, in this case, it is optimal for CAV $C$ to travel with constant acceleration and $t_1^* = 3.53s$. 
Proceeding to Phase II, the initial conditions are  $x_C(t_1^*)=101.92m,v_C(t_1^*)=34.67m/s,x_1(t_1^*)=128.99m,v_1(t_1^*)=28m/s,x_H(t_1^*)=101.92m,v_H(t_1^*)=26m/s$. We now solve problems (\ref{eq:hdv_ocp}), (\ref{eq:ibr_ocp_cavC}), and (\ref{eq:cav1_ocp}) followed by the IBR process described in Fig. \ref{fig:bilevel_diagram}. 
The weights for \textbf{OCP-HDV} in \eqref{eq:hdv_ocp_cost} are set to $\beta_u=0.9,\beta_v=0.1,\beta_s=0.1$. Additionally, we define $\mu=1,d=0$ in (\ref{eq:safety_cost}) when CAV $C$ is in the unsafe region of the HDV. We set the maximum number of iterations for the IBR process to $N=5$ (for the set of simulations results considered here, the process always converged within 5 iterations).

\begin{table}[pt]
\centering
\vspace*{-\baselineskip} \vspace{-1mm}
\caption{Vehicle $C$ Sample Results in Phase I.}
\label{tab:vehicleCSample}
\resizebox{\linewidth }{!}{%
\begin{tabular}{cccccc}
\toprule
    \diagbox[width=10em]{\textbf{{\ul OCPs}}}{\textbf{{\ul States}}} &
      \textbf{\begin{tabular}[c]{@{}c@{}}$X_C(t_0)$\\ {[}\textit{m}, \textit{m/s}{]}\end{tabular}} &
      \textbf{\begin{tabular}[c]{@{}c@{}}$X_1(t_0)$\\ {[}\textit{m}, \textit{m/s}{]}\end{tabular}} &
      \textbf{\begin{tabular}[c]{@{}c@{}}$X_H(t_0)$\\ {[}\textit{m}, \textit{m/s}{]}\end{tabular}} &
      \textbf{\begin{tabular}[c]{@{}c@{}}cost I\\ \end{tabular}} &
      \textbf{\begin{tabular}[c]{@{}c@{}}$t_1$\\ {[}\textit{s}{]}\end{tabular}} \\ \midrule
    (\ref{eq:OCP_cavC})  & [0,23] & [30,28] & [10,26] & Inf & Inf   \\ 
    (\ref{eq:OCP_cav1C})  & [0,23] & [30,28] & [10,26] & 2.99 & 4.18   \\ 
    (\ref{eq:constant_acc_profile})  & [0,23] & [30,28] & [10,26] & {\color{red}2.73} & {\color{red}3.53}   \\ \bottomrule
    \end{tabular}
    }%
\end{table}
\begin{figure*}[hpbt]
    \centering   
    \begin{subfigure}{0.4\linewidth}
    \centering 
      \includegraphics[angle=90, origin=c, width=\linewidth]{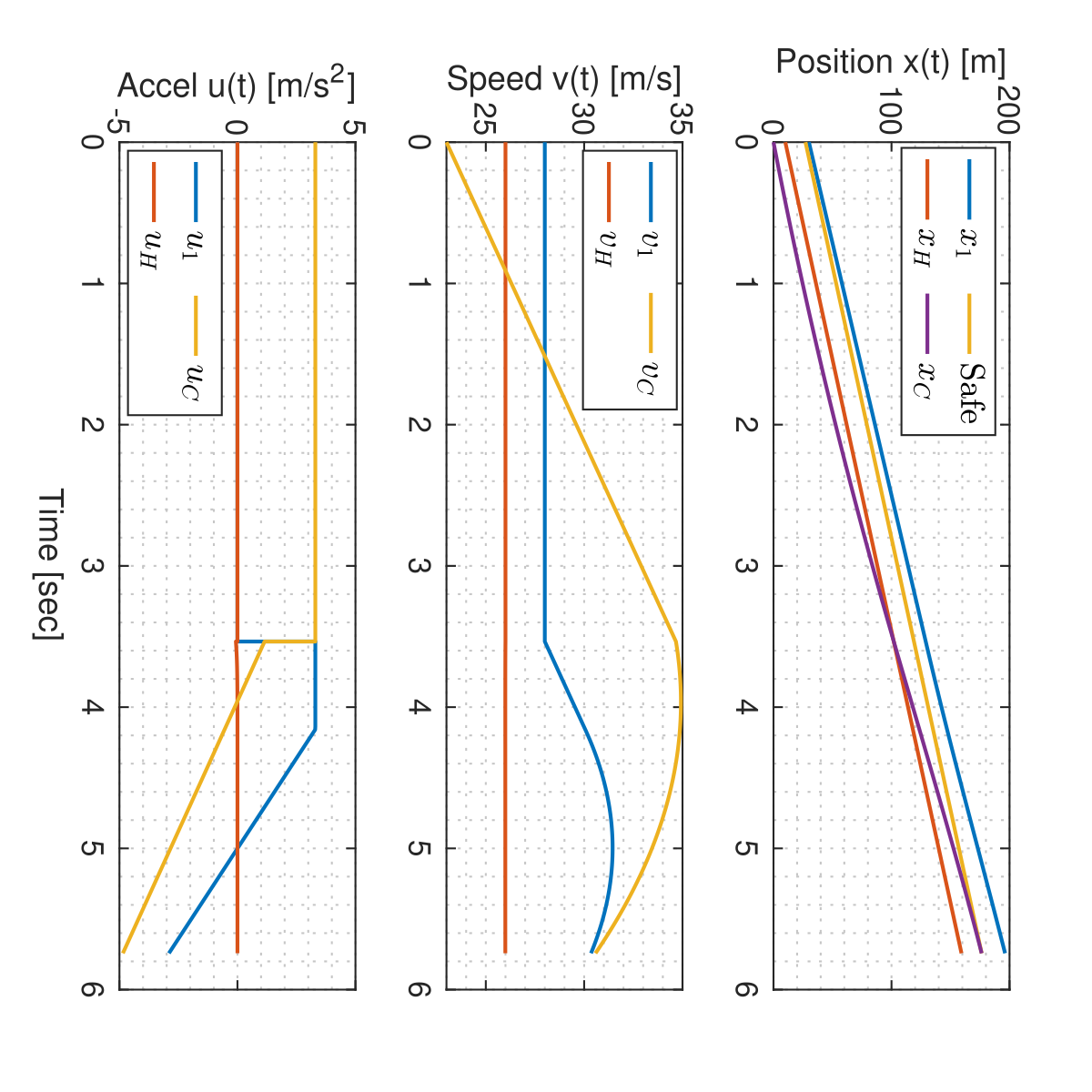}  
      \caption{\centering{Case 1: CAV C merges ahead of HDV}}
      \label{fig:phase1_constant_acceleration}
    \end{subfigure}   
    \begin{subfigure}{0.4\linewidth}
    \centering 
      \includegraphics[width=\linewidth]{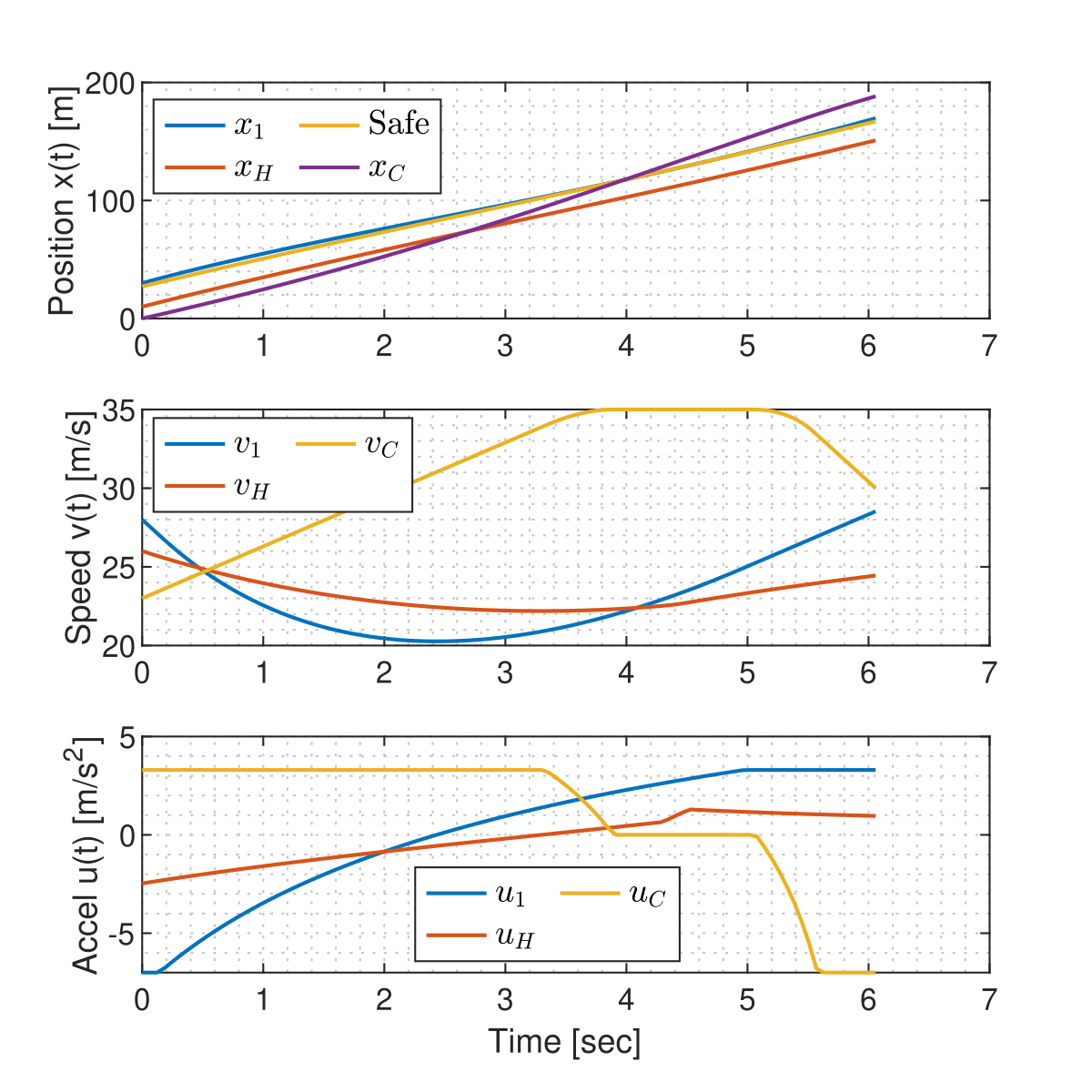}  
      \caption{\centering{Case 2: CAV C merges ahead of CAV 1  }}
      \label{fig:phase1_mixed_acceleration}
    \end{subfigure}
    \caption{Sample Optimal Trajectories for Vehicles $1,C,H$}
    \label{fig:CAV_C_phase2}  
\end{figure*}

\textbf{``Merge ahead of CAV 1'' policy.}
Similar to the previous case, we evaluate the cost of this policy by solving problem OCP (\ref{eq:cav1C_ocp}) using the same initial conditions as for the ``Merge ahead of HDV'' policy. 

\textbf{Computational cost.}
Even though an analytical solution of (\ref{eq:cav1C_ocp}) can be obtained, we considered here the ``worst case'' from a computational cost perspective and solved this problem numerically where our results took an average of 204 $ms$. We also note that the OCPs 
(\ref{eq:hdv_ocp}), (\ref{eq:ibr_ocp_cavC}), (\ref{eq:cav1_ocp}) each took an average 
of 50 $ms$ to solve.

\begin{table}[pt]
\centering
\vspace*{-\baselineskip} \vspace{-1mm}
\caption{Vehicle $C$ sample results for complete maneuvers}
\label{tab:vehicleCSample_total}
\resizebox{\linewidth }{!}{%
\begin{tabular}{ccccccc}
\toprule
    \diagbox[width=10em]{\textbf{{\ul Cases}}}{\textbf{{\ul States}}} &
      \textbf{\begin{tabular}[c]{@{}c@{}}$X_C(t_0)$\\ {[}\textit{m}, \textit{m/s}{]}\end{tabular}} &
      \textbf{\begin{tabular}[c]{@{}c@{}}$X_1(t_0)$\\ {[}\textit{m}, \textit{m/s}{]}\end{tabular}} &
      \textbf{\begin{tabular}[c]{@{}c@{}}$X_H(t_0)$\\ {[}\textit{m}, \textit{m/s}{]}\end{tabular}} &
      \textbf{\begin{tabular}[c]{@{}c@{}}cost\\ \end{tabular}} &
      \textbf{\begin{tabular}[c]{@{}c@{}}$t_f$\\ {[}\textit{s}{]}\end{tabular}} &
      \textbf{\begin{tabular}[c]{@{}c@{}}$dist$\\ {[}\textit{m}{]}\end{tabular}}\\ \midrule
    $C$ merges ahead of HDV  & [0,23] & [30,28] & [10,26] & {\color{red}4.47} & {\color{red}5.74} & 27.07  \\ 
    $C$ merges ahead of CAV 1  & [0,23] & [30,28] & [10,26] & 6.84 & 6.06  & 27.07 \\ 
    $C$ merges ahead of HDV  & [0,24] & [20,28] & [0,24] & 4.37 & 3.41 & 20  \\ 
    $C$ merges ahead of CAV 1  & [0,24] & [20,28] & [0,24] & {\color{red}3.99} & {\color{red}5.29} & 20  \\ \bottomrule
    \end{tabular}
    }%
\end{table}

The optimal trajectories over $t_0,t_f^*$ for all vehicles with the initial states defined in Table \ref{tab:vehicleCSample}  are shown in Fig. \ref{fig:CAV_C_phase2}. 
In Fig. \ref{fig:CAV_C_phase2}(a), when CAV $C$ merges ahead of the HDV, we observe that the purple line (CAV $C$) intersects the red line (HDV) at $t_1^*=3.53s$. The purple line overlaps the yellow line (safe distance ahead of HDV) around the terminal time $t_f^*=5.74s$. This illustrates the safety guarantee throughout the lane change maneuver. 
In Fig. \ref{fig:CAV_C_phase2}(b), when CAV $C$ merges ahead of CAV 1, the blue line (CAV 1) overlaps the yellow line, which indicates that the HDV is forced by CAV 1 to decelerate. At the terminal time, when CAV $C$ merges into the fast lane, the safety requirement is also guaranteed (the gap between the purple and blue lines at $t_f^*$). The total optimal costs and maneuver times 
for the two policies are summarized in Table \ref{tab:vehicleCSample_total}. 
It can be seen that the optimal policy depends on the distance $dist:=x_1(t_1)-x_H(t_1)$: as expected, when this distance is large, it is optimal for CAV $C$ to merge ahead of the HDV, otherwise it is optimal to merge ahead of CAV 1, in which case the HDV behavior is irrelevant to execute an optimal maneuver.

\subsection{Optimal CAV C Policy Criterion}

\begin{table*}[]
\centering
\caption{Cost and Disruption Comparison with $\beta_s=0.1$. Total Cost=$\sum\limits_{i=1,C,H}\text{Cost}\ i,$ CAVs=$\sum\limits_{i=1,C}\text{Cost}\ i$ }
\label{tab:cost_comparison}
\resizebox{0.9\linewidth }{!}{%
\begin{tabular}{cccccccccccc}
\toprule
    & \multicolumn{7}{c}{\textbf{Cost}} & \multicolumn{2}{c}{\textbf{HDV   disruption}} & \multicolumn{2}{c}{\textbf{Maneuver   time {[}s{]}}} \\ 
    \cmidrule{2-12} & \multicolumn{4}{c}{\textbf{C   merges ahead of H}} & \multicolumn{3}{c}{\textbf{C merges ahead of 1}} & \multicolumn{1}{c}{} & & \multicolumn{1}{c}{} & \\ 
    \cmidrule{2-8} \multirow{-3}{*}{\textbf{\begin{tabular}[c]{@{}c@{}}$dist$\\ 
    {[}m{]}\end{tabular}}} & \multicolumn{1}{c}{\textbf{Total}} & \multicolumn{1}{c}{\textbf{CAV1}} & \multicolumn{1}{c}{\textbf{CAVC}} & \multicolumn{1}{c}{\textbf{HDV}} & \multicolumn{1}{c}{\textbf{Total}} & \multicolumn{1}{c}{\textbf{CAVs}} & \textbf{HDV} & \multicolumn{1}{c}{\multirow{-2}{*}{\textbf{\begin{tabular}[c]{@{}c@{}}C merges \\
    ahead of 1\end{tabular}}}} & \multirow{-2}{*}{\textbf{\begin{tabular}[c]{@{}c@{}}C merges \\ ahead of H\end{tabular}}} & \multicolumn{1}{c}{\multirow{-2}{*}{\textbf{\begin{tabular}[c]{@{}c@{}}C merges \\ ahead of 1\end{tabular}}}} & \multirow{-2}{*}{\textbf{\begin{tabular}[c]{@{}c@{}}C merges \\ 
    ahead of H\end{tabular}}} \\ 
    \midrule
    20 & \multicolumn{1}{c}{4.37} & \multicolumn{1}{c}{0.07} & \multicolumn{1}{c}{3.04} & \multicolumn{1}{c}{1.26}          & \multicolumn{1}{c}{{\color[HTML]{FE0000} 3.99}} & \multicolumn{1}{c}{3.99} & 0.00 & \multicolumn{1}{c}{{\color[HTML]{3166FF} 0.07}} & {\color[HTML]{3166FF} 0.00} & \multicolumn{1}{c}{3.41} & 5.29 \\ 
    30 & \multicolumn{1}{c}{4.37} & \multicolumn{1}{c}{0.07} & \multicolumn{1}{c}{3.04} & \multicolumn{1}{c}{1.26}          & \multicolumn{1}{c}{{\color[HTML]{FE0000} 4.35}} & \multicolumn{1}{c}{4.35}          & 0.00          & \multicolumn{1}{c}{{\color[HTML]{3166FF} 0.07}} & {\color[HTML]{3166FF} 0.00} & \multicolumn{1}{c}{3.41} & 5.86 \\ 
    40 & \multicolumn{1}{c}{{\color[HTML]{FE0000} 4.37}} & \multicolumn{1}{c}{0.07} & \multicolumn{1}{c}{3.04} & \multicolumn{1}{c}{1.26} & \multicolumn{1}{c}{4.69} & \multicolumn{1}{c}{4.69} & 0.00 & \multicolumn{1}{c}{{\color[HTML]{3166FF} 0.07}} & {\color[HTML]{3166FF} 0.00} & \multicolumn{1}{c}{3.41} & 6.40 \\ 
    50 & \multicolumn{1}{c}{{\color[HTML]{FE0000} 4.37}} & \multicolumn{1}{c}{0.07} & \multicolumn{1}{c}{3.04} & \multicolumn{1}{c}{1.26} & \multicolumn{1}{c}{5.01} & \multicolumn{1}{c}{5.01} & 0.00 & \multicolumn{1}{c}{{\color[HTML]{3166FF} 0.07}} & {\color[HTML]{3166FF} 0.00} & \multicolumn{1}{c}{3.41} & 6.90 \\ 
    60 & \multicolumn{1}{c}{{\color[HTML]{FE0000} 4.37}} & \multicolumn{1}{c}{0.07} & \multicolumn{1}{c}{3.04} & \multicolumn{1}{c}{1.26} & \multicolumn{1}{c}{5.32} & \multicolumn{1}{c}{5.32} & 0.00 & \multicolumn{1}{c}{{\color[HTML]{3166FF} 0.07}} & {\color[HTML]{3166FF} 0.00} & \multicolumn{1}{c}{3.41} & 7.39 \\ 
    70 & \multicolumn{1}{c}{{\color[HTML]{FE0000} 4.37}} & \multicolumn{1}{c}{0.07}          & \multicolumn{1}{c}{3.04}          & \multicolumn{1}{c}{1.26}          & \multicolumn{1}{c}{5.62}                        & \multicolumn{1}{c}{5.62}          & 0.00          & \multicolumn{1}{c}{{\color[HTML]{3166FF} 0.07}}                                                               & {\color[HTML]{3166FF} 0.00}                                                               & \multicolumn{1}{c}{3.41}                                                                                      & 7.85                                                                                      \\  
80                                                                                    & \multicolumn{1}{c}{{\color[HTML]{FE0000} 4.37}} & \multicolumn{1}{c}{0.07}          & \multicolumn{1}{c}{3.04}          & \multicolumn{1}{c}{1.26}          & \multicolumn{1}{c}{5.91}                        & \multicolumn{1}{c}{5.91}          & 0.00          & \multicolumn{1}{c}{{\color[HTML]{3166FF} 0.07}}                                                               & {\color[HTML]{3166FF} 0.00}                                                               & \multicolumn{1}{c}{3.41}                                                                                      & 8.29                                                                                      \\ 
90                                                                                    & \multicolumn{1}{c}{{\color[HTML]{FE0000} 4.37}} & \multicolumn{1}{c}{0.07}          & \multicolumn{1}{c}{3.04}          & \multicolumn{1}{c}{1.26}          & \multicolumn{1}{c}{6.19}                        & \multicolumn{1}{c}{6.19}          & 0.00          & \multicolumn{1}{c}{{\color[HTML]{3166FF} 0.07}}                                                               & {\color[HTML]{3166FF} 0.00}                                                               & \multicolumn{1}{c}{3.41}                                                                                      & 8.72                                                                                      \\ 
100                                                                                   & \multicolumn{1}{c}{{\color[HTML]{FE0000} 4.37}} & \multicolumn{1}{c}{0.07}          & \multicolumn{1}{c}{3.04}          & \multicolumn{1}{c}{1.26}          & \multicolumn{1}{c}{6.46}                        & \multicolumn{1}{c}{6.46}          & 0.00          & \multicolumn{1}{c}{{\color[HTML]{3166FF} 0.07}}                                                               & {\color[HTML]{3166FF} 0.00}                                                               & \multicolumn{1}{c}{3.41}                                                                                      & 9.14                                                                                      \\ \bottomrule
\end{tabular}

}
\end{table*}
\begin{figure*}[hpbt]
    \centering   
    \begin{subfigure}{0.45\linewidth}
    \centering 
      \includegraphics[width=\linewidth]{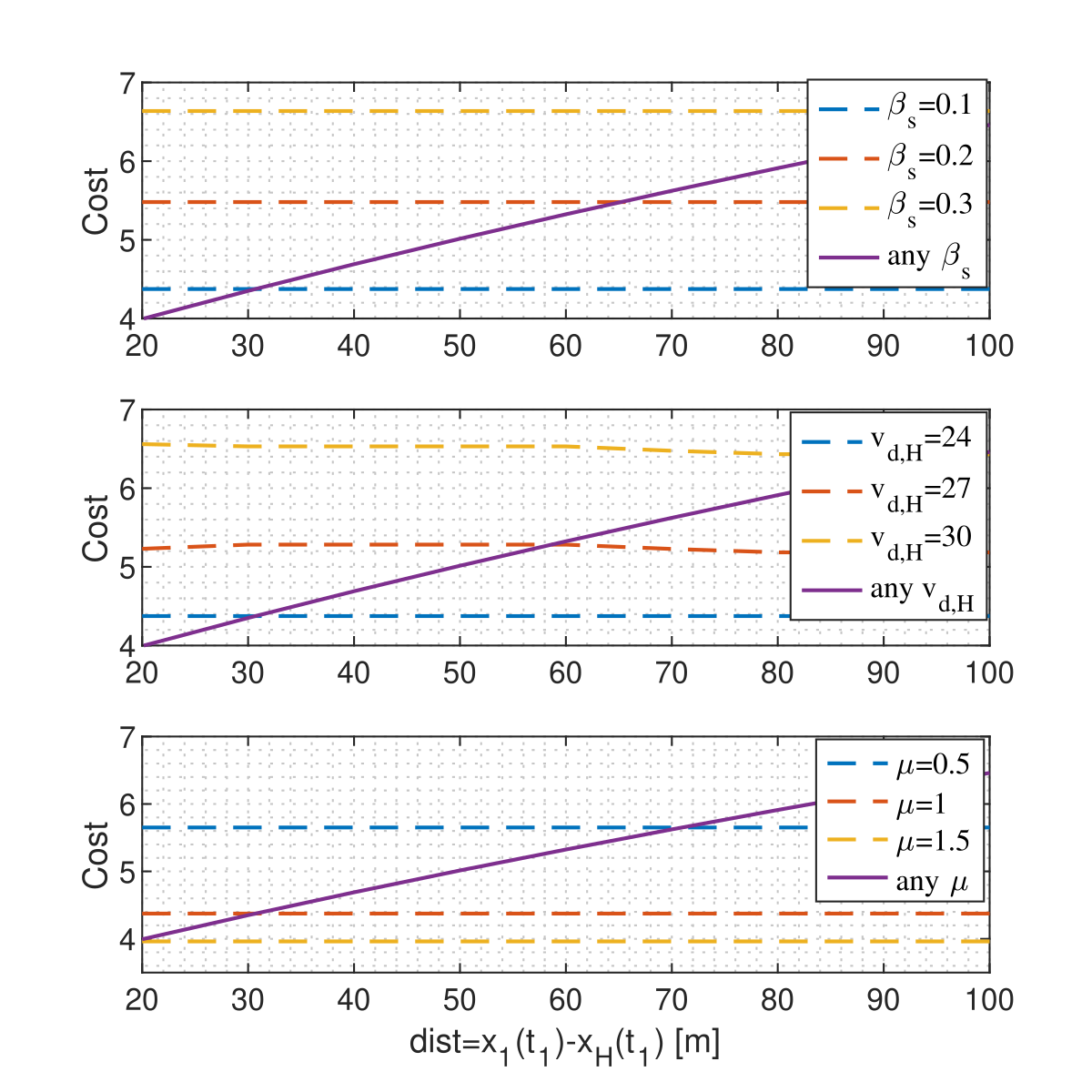}  
      \caption{\centering{Total cost under different parameters $\beta_s,v_{d,H},\mu$}}
      \label{fig:phase2_cost}
    \end{subfigure}   
    \begin{subfigure}{0.45\linewidth}
    \centering 
      \includegraphics[width=\linewidth]{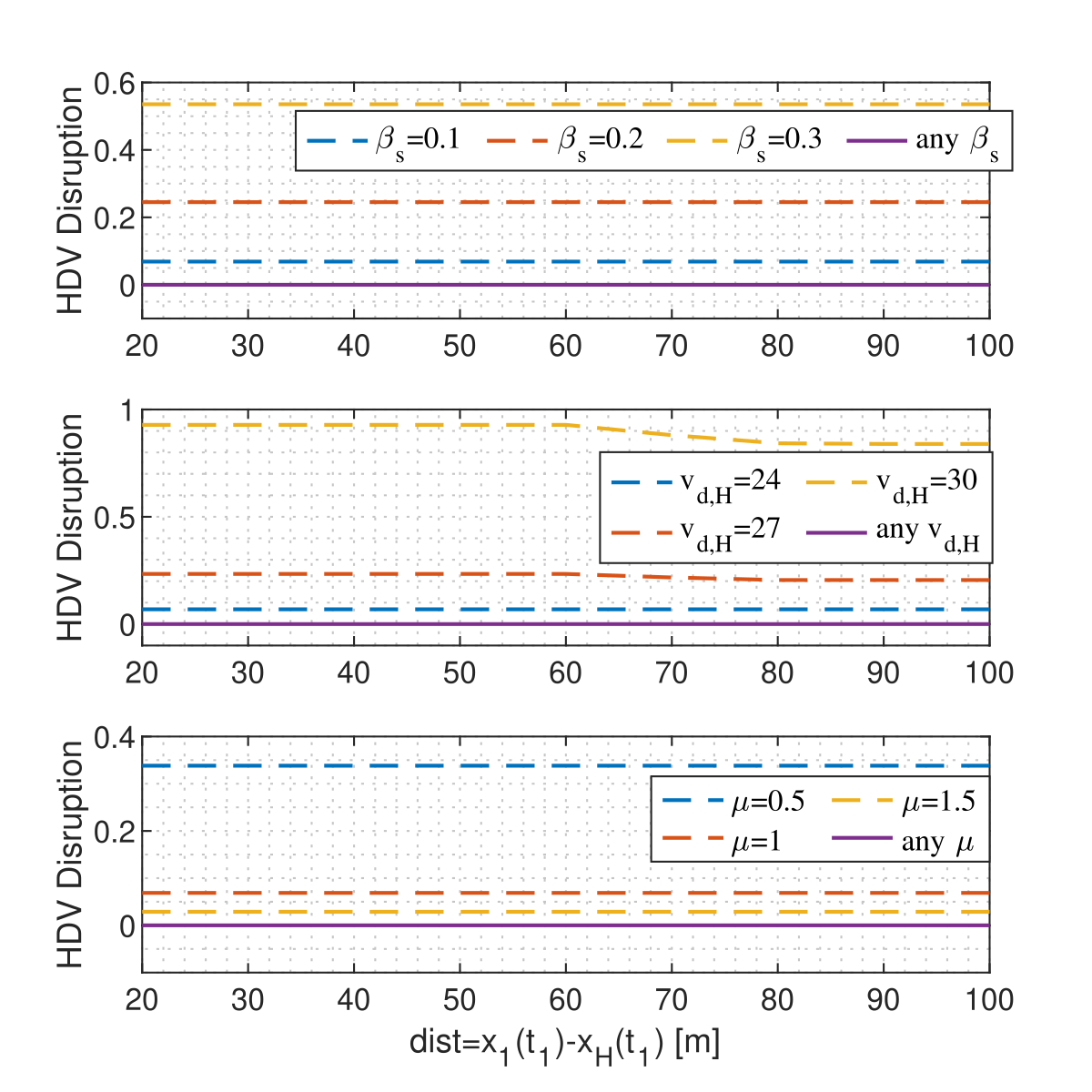}  
      \caption{\centering{Disruption of HDV under different parameters $\beta_s,v_{d,H},\mu$}}
      \label{fig:phase2_disruption}
    \end{subfigure}
    \caption{Cost and disruption comparison}
    \label{fig:phase2_cost_disruption}  
\end{figure*}

In this section, we seek a ``\textit{simple-to-detect}'' criterion for CAV $C$ to determine an optimal policy while also taking into account the \textit{traffic disruption} as defined in (\ref{eq:totaldisruption}). Moreover, we explore the effect of various  parameters that characterize the HDV behavior on policy determination. Specifically,
the weight $\beta_s$ on the safety component of the HDV cost function in (\ref{eq:hdv_ocp}); the HDV's desired speed $v_{d,H}$; and the parameter $\mu$ in the safety cost (\ref{eq:safety_cost}). 

In what follows, we omit Phase I and focus on the phase of the maneuver that includes the interaction between the HDV and the two CAVs.
The initial speeds of vehicles are set as $v_1(t_1)=28m/s,\, v_C(t_1)=v_H(t_1)=24m/s,$ and, for simplicity, the initial positions of CAV $C$ and $H$ are $x_H(t_1)=x_C(t_1)=0$. The cost weights are set as $\alpha_u=0.2$, $\alpha_v=0.8$, $\beta_u=0.9$, $\beta_v=0.1$. 
Fixing $\beta_s=0.1$, $\mu=1$,
Table \ref{tab:cost_comparison} summarizes the vehicle costs, disruption to the HDV, and maneuver time under each of the two CAV $C$ policies and for different values of $dist$. The total cost is defined as the sum of each of the costs from every vehicle (CAV 1, CAV2, and HDV).
Observe that when $dist$ increases from $20m$ to $40m$, the optimal policy for CAV $C$ switches from ``merge ahead of CAV 1'' to ``merge ahead of HDV'' if $C$ aims to complete a minimal cost maneuver without taking disruption into account. Note that the maneuver time for CAV $C$ in the ``merge ahead of HDV'' policy remains constant, which means 
$dist$ does not affect the feasibility and optimality of the maneuver under the current settings. The unchanged maneuver time corresponds to a fixed optimal policy for $C$ under different $dist$. Therefore, the cost for each vehicle $i$ also remains constant.
The total cost for CAV $C$ to ``merge ahead of CAV 1'' is monotonically increasing with respect to $dist$, since completing the maneuver incurs a higher effort by the CAVs with a larger $x_1(t_1)-x_C(t_1)$. As for the disruption to the HDV, observe that ``merge ahead of HDV'' leads to some small HDV disruption while the policy ``merge ahead of CAV 1'' causes no disruption at all. 

The total cost and disruption comparison of CAV $C$ merges ahead of HDV and CAV $C$ merges ahead of CAV 1 with respect to $dist$ under different $\beta_s,v_{d,H},\mu$ are illustrated in Fig. \ref{fig:phase2_cost_disruption}, in which the dashed lines represent the ``merge ahead of HDV'' policy, while the straight lines represent the ``merge ahead of 1'' policy. In Fig. \ref{fig:phase2_cost_disruption}(a), when increasing $\beta_s$, the total cost will also increase since a larger $\beta_s$ corresponds to a more conservative driver. Once it is feasible for CAV $C$ to change lanes under a safety weight, continuing increasing $\beta_s$ will cause redundant braking while consuming more energy. $v_{d,H}$ is the desired speed of HDV, and a higher $v_{d,H}$ means both $C$ and HDV have to accelerate more to complete the maneuver safely and to achieve the desired speed. The parameter $\mu$ represents how the HDV defines its safe region. A larger $\mu$ corresponds to a smaller safe region. Hence, decreasing $\mu$ means extending the safe region. Similarly, by increasing $\mu$ the cost for the OCP will also increase. Since all the parameters $\beta_s,v_{H,d},\mu$ are contained in HDV's problem, which is irrelevant to the CAVs if $C$ chooses to merge ahead of 1. The total cost is independent of all the above parameters and monotonically increases when increasing the $dist$. The reason is that a larger $dist$ means that CAV $C$ needs to spend more time and energy to merge ahead of CAV 1 while forcing CAV 1 to decelerate harder to decrease $C$'s travel time. Thus, a potential deceleration can lead to a higher total cost. It is also worth noting that the dashed line intersects with other lines. The intersections provide thresholds for the relative distance between vehicles at the initial time such that CAV $C$ can choose a better strategy to complete the lane change maneuvers with minimal cost. 

As for the HDV disruption in Fig. \ref{fig:phase2_cost_disruption}(b), ``merge ahead of HDV" (dashed lines) leads to a higher disruption than ``merge ahead of CAV 1" (straight line). The reason is that for $C$ to ``merge ahead of HDV"  requires extra deceleration than ``merge ahead of CAV 1". Thus, when increasing $\beta_s$, the HDV has to brake harder to guarantee safety with respect to CAV C,  causing a higher disruption. If the HDV aims to reach a higher desired speed $v_{d,H}$, the disruption will increase according to (\ref{eq:totaldisruption}). With a similar analysis, a larger safe region represents a more conservative driver, so the cost will increase as well. Note that when CAV $C$ merges ahead of 1, the response from HDV becomes irrelevant. No matter how aggressive or conservative the HDV is, the disruption to HDV will be unchanged. For the ``merge ahead of 1" case, CAV 1 can recover to a higher terminal speed of $30m/s$ optimally while minimizing the disruption to HDV. 

    

\subsection{Comparison with 
Human-Driven Vehicles}


We use the standard car-following models in SUMO to simulate lane change maneuvers involving HDVs only (baseline). In this case, vehicles 1 and $H$ are defined as $C$'s immediate left leader and left follower, respectively at the time  when $C$ decides to change its lane. In the baseline case, it is worth noting that no HDV chooses to merge ahead of vehicle 1 because there is no cooperation between vehicle 1 and $C$. In this case, $C$ has to accelerate harder and consume energy to overtake vehicle 1 and perform the maneuver with collision risks. The comparison of the costs and disruptions for the $100\%$ HDV case are shown in Table. \ref{tab:baseline_comparison}. With the given initial states,  the ``merge ahead of $H$" policy provides a lower cost and shorter maneuver time than the ``merge ahead of $1$" policy. However, ``merge ahead of 1" still has 0 disruptions to vehicle H, so as to the following fast lane traffic. Our policies can save more than $80\%$ in cost and almost eliminate the disruption to the fast lane traffic. 

\begin{table}[]
\caption{Baseline Results Comparison}
\label{tab:baseline_comparison}
\resizebox{\linewidth }{!}{%
\begin{tabular}{cccc}
\toprule
\textbf{Scenarios }          & \textbf{TotalCost} & \textbf{HDV disruption}  & \textbf{Maneuver Time {[}s{]}} \\ \midrule
Baseline            & 22.371    & 678.05               & 7.3825                \\ 
C merges ahead of H & 2.847     & 0.165               & 2.923                 \\
C merges ahead of 1 & 3.917     & 0                   & 6.388                 \\ \bottomrule
\end{tabular}}
\end{table}

\section{CONCLUSIONS AND FUTURE WORK}
\label{secV:Conclusions}
We have developed optimal control strategies for a CAV to complete lane change maneuvers while minimizing the travel time, energy and speed disruption to the traffic flow in mixed traffic. Vehicle interactions and cooperation have been considered to help optimally perform the maneuver. The simulation results show the effectiveness of the proposed controllers and provide a criterion for $C$ to always choose a policy with minimal cost. The limitation of this work is to assume the objectives and dynamics of HDVs are known to CAVs, and that there are no disturbances or uncertainties in the network, which are difficult to achieve in the real world. Our ongoing works aim to figure out the characteristics of different drivers in real-time and extend the lane change maneuvers in the mixed traffic scenario from a single vehicle to multi vehicles. Besides, how to increase the probability of HDVs cooperating with CAVs is also a promising way to explore.

\addtolength{\textheight}{-12cm}   





\bibliographystyle{IEEEtran}

\bibliography{cmp}

\end{document}